\documentclass[usegraphicx,useAMS,usenatbib]{mn2e}
\def\kms{\hbox{km s$^{-1}$}}
\def\hii{\hbox{H\,{\sc ii}}}
\def\oiii{\hbox{[O\,{\sc iii}]}}
\def\fevii{\hbox{[Fe\,{\sc vii}]}}
\def\fex{\hbox{[Fe\,{\sc x}]}}
\def\neii{\hbox{[Ne\,{\sc ii}]}}
\def\sivi{\hbox{[Si\,{\sc vi}]}}
\def\sivii{\hbox{[Si\,{\sc vii}]}}
\begin{document}
%%%%%%%%%%%%%%%%%%%%%%%%%%%%%%%%%%%%%%%%%%%%%%%%

\title[Seyfert galaxies at 11 and 18 micron]{VLT diffraction-limited 
imaging at 11 and 18 $\mu$m of the nearest active galactic nuclei \thanks{Based on observations collected at the
European Southern Observatory, VLT programs 076.B-0599, 077.B-0728}}
\author[Reunanen et al.]{J. Reunanen$^1$,
M. A. Prieto$^2$,
R. Siebenmorgen$^3$\\
$^{1}$Tuorla Observatory, University of Turku, FI-21500 Piikki\"o, Finland; reunanen@ftml.net\\
$^{2}$Instituto de Astrof\'isica de Canarias, C/ V\'ia L\'actea, 38205 La Laguna, Spain; aprieto@iac.es \\
$^{3}$European Southern Observatory, Karl-Schwarzschild-Strasse 2, D-85748 Garching bei M\"unchen, Germany}

\date{}

\pagerange{\pageref{firstpage}--\pageref{lastpage}} \pubyear{2009}

\maketitle

\label{firstpage}

\begin{abstract}
  Mid-infrared (IR)  imaging at resolutions of 300 mas of the central kpc
  region of  13 nearby, well-known active galaxies is presented.  
The bulk of the mid-IR emission is concentrated on an unresolved central 
source within a size of less   than 5 to 130 pc, depending on the object distance.  Further
  resolved emission is detected in 70\% of the sample in the form of
  circumnuclear star-forming rings or diffuse nuclear extended
  emission. In the three cases with circumnuclear star formation, the
  stellar contribution is at least as important as that
  of the AGN. In those with extended nuclear emission -- a third of
  the sample -- this emission represents a few per cent of the total
  measured; however, this contribution may be underestimated because of 
the chopped nature of these  observations.
  This extended emission is generally collimated in a
  preferential direction often coinciding with that of the extended
  ionized gas or the jet. In M87 and Cen A, where the emission  extends 
along their respective jets, the emission is presumably synchrotron. In Circinus, 
NGC 1386 and NGC 3783, it can be reconciled with 
  thermal emission from dust heated at about 100 K by the active nucleus.
    
  In all cases, the nuclear fluxes measured  at 11.8 $\mu$m and 18.7 $\mu$m 
represent a minor contribution of the  flux levels measured by large aperture 
\textit{IRAS} data at the nearest energy bands of 12 and 25 $\mu$m. 
This contribution  ranges from   30\% to less than 10\%.  In only three 
cases do the active galactic nucleus (AGN) fluxes agree with
  \textit{IRAS} to within a factor of 2. In the AGNs with strong circumnuclear star formation, this
  component can well account for most of the \textit{IRAS} flux measured in
  these objects. But in all other cases, either a low surface brightness
  component extending over galactic scales or strong extra-nuclear IR
  sources -- e.g. {\hii} regions in spiral arms -- have to be the main
  source of the \textit{IRAS} emission. In either case, the
  contribution of these components dwarfs that of the AGN at mid-IR
  wavelengths.

\end{abstract}
 

%%%%%%%%%%%%%%%%%%%%%%%%%%%%%%%%%%%%%%%%%%%%%%%%%%%%%%%%%%%%%%%%%%%%%%%%%
\section{Introduction}

With the advent of interferometry and adaptive optics techniques in
the infrared (IR) in large ground-based telescopes, the central
regions of active galactic nuclei (AGN) are being studied in
ever-increasing detail. The best details have so far been seen on the
brightest and nearest AGN observed with the Very Large Telescope (
VLT). Here is a list. The nucleus of NGC 1068 is resolved into a
parsec-scale disc in near-IR (Weigelt et al.\ 2004) and mid-IR (Jaffe
et al.\ 2004) interferometry observations.  That of Circinus is also
resolved into a parsec-scale disc-like structure perpendicular to the
ionization cone in adaptive optics images in the near IR (Prieto et
al.\ 2004), the results being further confirmed by interferometry in
the mid-IR (Tristram et al.\ 2008). In both cases, the resolved
nuclear structure fits within the characteristics of a parsec-scale
torus.  The nucleus of Cen A, however, is so far unresolved down to
scales of less than a parsec in adaptive optics images in the near IR
(Haring-Neumayer et al. 2006) and interferometry in the mid-IR
(Meisenheimer et al.\ 2007).  Other bright AGN observed with adaptive
optics in the near IR show an unresolved nucleus, so far, down to
scales of tens of parsecs (Prieto et al.\ 2007), and from those that
could be targeted with interferometry in the mid-IR, the available
data indicate a resolved nuclear structure, $\sim 2$ pc in size, and
NGC 4151, and possible resolved emission in NGC 1365, NGC 3783 and NGC
7469 (Tristram et al. 2009).  The torus structure may be larger than a
few parsecs at mid-IR wavelengths, where dust heated by the AGN to
blackbody equivalent temperatures of 100--300 K may exist at larger
radii. One of the problems is that this outermost region may be fully
resolved with interferometry and thus will escape detection.

This work presents difraction-limited VLT observations at 11.8 $\mu$m
and 18.7 $\mu$m of a sample of the nearest and brigthest AGN
accessible from Paranal Observatory. The resolutions achieved are a
factor 10 lower than those provided by VLT-MIDI interferometry, but
are more sensitive to possible extended structures around the nucleus
(e.g. Perlman et al. 2001; Siebenmorgen et al. 2008); at the very
least, they allow for setting an upper limit to the outer radius of
the torus. At the higher resolution achieved in these observations
(FWHM = 0.3 arcsecs) this upper-limit radius in the sample galaxies is
35 pc on average (ranging from 5 to 130 pc).

This paper is part of an major project focusing on the study of the
central few parsecs of the nearest AGN at optical, IR and radio
wavelengths, using the highest spatial resolution data available
today. The sample galaxies have been extensively studied across all
ranges of the electromagnetic spectrum. For all these objects VLT
subarcsec resolution observations, by means of adaptive optics in the
1 to 5 $\mu$m range \citep{Prieto2009}, and at 11.9 and 18.7 $\mu$m
(this work) were collected.  These data are complemented with
\textit{HST} optical information available for all sources, and VLA
and/or ATCA data at subarcsec resolution \citep{Orienti2009}.  At the
distance of the sample galaxies, the spatial scales at which the
nuclear regions are studied range from a few pc in the optical,
near-IR and radio to several tens of parsecs in the mid-IR.  A
comprehensive study of the spectral energy distribution of these AGN
using all the available high spatial resolution data is presented in
\citet{Prieto2009}.

Throughout this paper, $H_0$ = 71 {\kms} Mpc$^{-1}$ is used.
\section{Observations and data reduction}

\label{reduction}

\begin{table*}
  \caption{The AGN sample observed with VISIR. The distance, $D$, was derived 
    from the velocity $v_{3K}$ provided in NED, except for NGC~1386, NGC~5128 and Circinus, which are from
    Madore et al. (1999), \citet{Ferrarese2007} and \citet{Freeman1977}, 
    respectively. Galaxy morphology and  classification are from NED. The 12 and 
25 $\mu$m IRAS fluxes are from \citep{IRASF}. The last two columns give  total on-source 
integration time.}
\label{table_sample}
\begin{tabular}{@{}lcccllcccccc}
\hline 
Galaxy &  $z$  & $D$ & Scale &Host type & AGN &12$\mu$m & 25$\mu$m& $t_\mathrm{11.8}$ & $t_\mathrm{18.7}$ &\multicolumn{2}{c}{Observing date}\\
       &       & Mpc & pc/arcsec&          &         & Jy & Jy&  s  &  s & 11.8 & 18.7 \\
\hline
NGC 1097    & 0.004240& 15.1&~73 & SBb       & Sy1  & 1.985 & 5.509 &~660&1000 & 2006-09-04 & 2006-09-05 \\
NGC 1386    & 0.002895& 18.6&~90 & SB0       & Sy2  & 0.493 & 1.433 &~600&1000 & 2006-08-17 & 2006-08-16 \\
NGC 1566    & 0.005017& 20.5&~99 & SABbc     & Sy1  & 0.831 & 1.219 &~600&1120 & 2006-09-05 & 2006-09-05 \\
ESO 434-G040& 0.008486& 39.2&190 & SA0       & Sy2  &       &       &~600&1800 & 2006-04-28 & 2005-12-18 \\
& & & & & & & & & & & 2006-03-13 \\
Mrk 1239    & 0.019927& 86.6&420 & E-S0      & Sy1.5& 0.65~ & 1.141 &~600&1000 & 2006-04-28 & 2006-05-16 \\
NGC 3783    & 0.009730& 44.3&215 & SBa       & Sy1  & 0.840 & 2.492 &~600&600  & 2006-04-29 & 2006-03-13 \\
M87         & 0.004360& 22.4&109 & E0--1 pec & Sy   & 0.231 &$<$0.25&3560&1870 & 2006-03-09 & 2006-03-09 \\
& & & & & & & & & & 2006-05-17 \\
Sombrero    & 0.003416& 18.7&~91 & SAa       & Sy1.9& 0.386 & 0.497 &~800&...  & 2006-03-22 & ... \\
NGC 5128    & 0.001825& ~3.4&~17 & S0 pec    & Sy2  & 13.26 & 17.26 &~300&~750 & 2006-03-15 & 2006-03-22 \\
NGC 5506    & 0.006181& 29.0&141 & Sa pec    & Sy1.9& 1.282 & 3.638 &~600&1000 & 2006-06-06 & 2006-06-06 \\
Circinus    & 0.001448& ~4.2&~20 & SAb       & Sy2  & 18.8  & 68.44 &~500&1000 & 2006-06-05 & 2006-03-27 \\
NGC 7469    & 0.016317& 61.9&300 & SABa      & Sy1.2& 1.348 & 5.789 &~600&1020 & 2006-07-14 & 2006-08-06 \\
NGC 7582    & 0.005254& 18.3&~89 & SBab      & Sy2  & 1.620 & 6.436 &~750&1000 & 2006-08-06 & 2006-06-16 \\
\hline
\end{tabular}
\end{table*}

The basic properties of the AGN sample are shown in
Table~\ref{table_sample}.  The galaxies were observed with the
$256\times256$ pixel VLT Imager and Spectrometer for mid Infrared
(VISIR; \citealt{VISIR_ref} with the pixel scale of 0.075
arcsec/pixel. The filters were selected to be free from polyaromatic
hydrocarbon (PAH) features: a filter centred at 11.88 $\mu$m and with
half bandwidth of 0.37$\mu$m, and a second one centred at 18.72
$\mu$m and half bandwidth of  0.88 $\mu$m were used. In most cases the nodding
direction was taken perpendicular to the chopping direction, keeping
the four resulting beams inside the VISIR field of view, which
effectively limits the total field of view to
$\sim9\times9$ arcsec. Identical distances, typically 8 or
10 arcsec, were used for nodding and chopping. A few galaxies, which
were known \textit{a priori} to harbour extended emission --
star-forming clusters or rings -- were observed in parallel nodding
mode, providing a field of view of 19.2 $\times$ 19.2 arcsec. Each
science observation was immediately followed by a standard star, which
was used for both photometric calibration and for point spread
function (PSF) control.  The accuracy of the absolute flux calibration
is limited by the uncertainties in the flux calibration of the mid-IR
standard stars and is estimated to be $\sim$10\%. Errors given in
the later sections and tables are statistical.

Data reduction was done with the ESO pipeline, which was further
modified by us to remove the striping pattern of the detector and to
improve the centring of the chopped frames.  Using service mode for
the observations guaranteed both reaching the diffraction limit and
acceptable levels for striping. As the intensity of striping depends
roughly on the brightness of the sky and its variability, observations
with strong striping were rejected by the telescope staff and repeated
at a later time. Low-level striping was easy to remove by fitting and
subtracting a low-order polynomial along image rows with suitable
pixel rejection criteria to guarantee that the observed galaxy or star is
not included in the fit.

The data reduction procedure consists of simply shifting and stacking
 chopped/nodded frames. To provide the most accurate combination of
frames for bright sources, the shortest individual exposures at each
chopping position -- typically an integration time of a few seconds -- were
combined by using centroiding on the source itself (marked with a $^*$
in Table \ref{table_flux}). For fainter sources centroiding was done
on the combined nodding/chopping exposure -- typically representing
around 1 minute of integration time (marked with a $^{**}$ in Table
\ref{table_flux}). A few sources are too faint at 11.8 $\mu$m for any
centroiding and we were forced to rely on telescope guiding for
combining each beam. This is typically also the case at 18.7
$\mu$m. The same combination scheme was always used for both the
galaxy and its associated PSF standard. Finally, the four beams were
combined.

The accuracy of the absolute flux calibration is limited by the
uncertainties in the flux calibration of the mid-IR standard stars and
is estimated to be 5 - 10\%. Flux errors given in the tables are
statistical.  When available, we compared our VISIR photometry with
equivalent measurements reported in the literature. The agreement is
in general good within the errors, but some discrepancies exist.  The
result of this comparison is discussed on a case basis at each object
section.
%%%%%%%%%%%%%%%%%%%%%%%%%%%%%%%%%%%%%%%%%%%%%%%%%%%%%%%%%%%%%%%%%%%%%%%%%
\section{Analysis}

\subsection{Quality of the PSF\label{PSF}}

The full width at half maximum (FWHM) resolution achieved in the
standard star images is typically 0.31--0.35 arcsec at 11.8~$\mu$m and
0.49--0.53 arcsec at 18.7~$\mu$m (Table \ref{table_size}). The
observations of the PSF stars often show an elongation, predominantly
along the nodding direction. The reason for the elongation is unclear:
\citet{Tokovinin2007} suggests that it may be related to the tilt
anisoplanatism between the guide star and the object. The direction of
elongation does not depend on the rotation angle of the instrument and
is therefore not caused by the support structure of the secondary
mirror.  Table \ref{table_size} gives the residual ellipticity of the
images $e= 0.10-0.24$ at 11.8 $\mu$m ($e=0.15$ on average) and
slightly lower $e= 0.03-0.15$ ($e=0.10$ on average) at 18.7 $\mu$m.

\begin{figure*}
  \includegraphics[height=5.20cm]{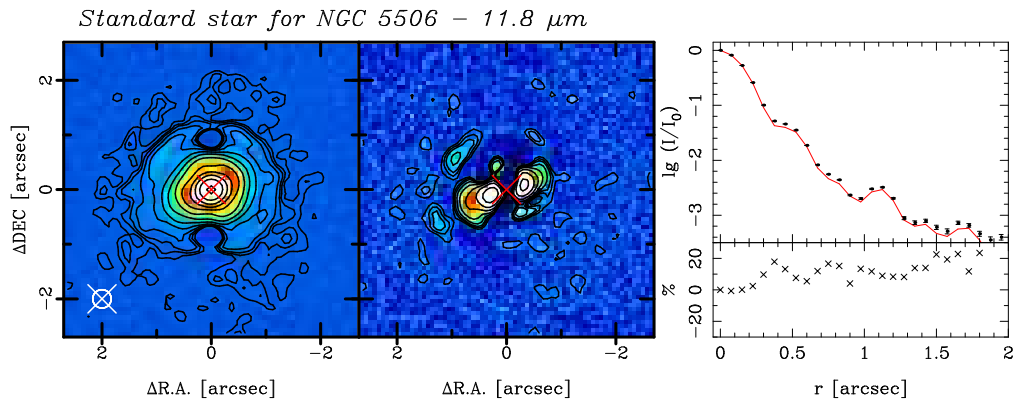}

  \caption{The 11.8 $\mu$m PSF standard of NGC 5506.  \textit{Left}:
    observed emission. \textit{Middle}: the residual after
    subtracting the average PSF. \textit{upper Right}: radial
    profiles in logarithmic flux units normalized at $r=0$ for the
    standard star plotted with errorbars and the average PSF standard
    (\textit{red}line). The circle on the lower left
    corner notes the FWHM size, while the the size of the cross at the
    nucleus indicates the size of the first diffraction ring
    (1.22$\lambda\, D^{-1}$). \textit{Lower right} the residual (observed
    - average PSF) in \%s is shown. North is up, and east is to the left, and the
    distances are given in arcseconds. The lowest contours are at a
    3, 5, 7, 11, 19 level. 1$\sigma$ is 23 mJy arcsec$^{-2}$.\label{psf_n5506}}
\end{figure*}

In order to investigate this intrinsic, variable elongation and its
effect on the analysis of the science data further, an average of all
standard star observations was taken and subtracted this from the
individual standard star observation. A typical residual was a pair of
central symmetric lobes with varying position angle
(Fig. \ref{psf_n5506}). The same residual is often present when
comparing the galaxy observation with its associated PSF star
(e.g. Fig. \ref{fig_5506_circ}) and is therefore taken to be an
artefact in the present analysis.

The FWHM of the PSF stars is rather stable and similar to that
measured at the galaxies nucleus. The standard deviation of the FWHM
measurements in the PSF star frames is $\sigma=0.014$ arcsec at 11.8
$\mu$m and $\sigma=0.013$ arcsec at 18.7 $\mu$m. In two cases the
nucleus FWHM is significantly higher than that of the corresponding
PSF star, both in the 18.7 $\mu$m images: NGC 7582, where the extended
emission from the star-forming ring affects the measurement of the
core width, and NGC 1386, which presents the strongest extended
nuclear emission in the sample.

\subsection{Extended emission}
In all cases, the most conspicuous source at these wavelengths is the
nucleus. The detection of low surface brightness emission around the
nucleus is limited by the chopping technique used in ground mid-IR
observations. Nevertheless, in some cases extended nuclear emission
and circumnuclear star-forming regions are strong enough to be
detected. To search for extended emission, both a PSF-subtraction and
a radial profile analysis were used.  In the first case, the PSF star
was subtracted from its associated-in-time galaxy observation after
normalizing it to the galaxy peak at the nucleus.  The residual map
thus has zero flux at the very centre, and possible differences in the
shapes of the PSF star and galaxy are then visible as negative or
positive fluxes.  In some cases, these residuals take the form of two
symmetric central blobs and should be ignored (see Sect. 3.1). In
addition, radial profiles were extracted by azimuthally averaging the
emission about the nucleus. This analysis is more sensitive to the
presence of extended emission due to the accumulated higher S/N and is
also less affected by PSF variations. Therefore radial profiles are
used as primary indicators for the presence of extended emission.

\subsection{Nuclear photometry\label{sec_nuc}}
Nuclear point-like fluxes were derived from both the radial profile
analysis and direct aperture photometry (Table \ref{table_flux}). In
cases where the emission is clearly extended, a nuclear point-like
flux (Table 3) is derived from a fit to the radial profile. The fit
has two components: a disc profile ($I=I_0\,\exp^{-r/r_0}$) convolved
with the associated PSF star representing the extended component, and
the profile of the PSF star representing the unresolved nucleus. For
some galaxies, the associated PSF star was too faint at the wing
levels. In that case, a combined PSF profile was created by replacing
the region beyond $\sim$1 arcsec with the profile of the brightest PSF
star in the sample, HD~2261 (35 Jy at 11.8 $\mu$m and 15 Jy at 18.7
$\mu$m).

In galaxies with weak extended emission, nuclear aperture photometry
was used instead.  The optimal aperture radius was determined by
comparing the galaxy radial profile with that of the PSF star. The
radius at which the galaxy radial profile began to diverge from that
of the PSF was selected and a core flux derived. The point-like flux
in Table 3 is the core value multiplied by a correction factor -- to
account for the additional unresolved nuclear flux contained in the
PSF wings.
 
Finally, in pure unresolved nuclear sources, fluxes are directly those
measured in an aperture containing all the observed emission. We note,
however, that in these cases emission from the nuclear PSF wings is
weak or undetected. Following the same procedure as before, a
correction factor -- derived from the PSF star radial profile -- to
account for the additional flux in the nuclear PSF wings was applied
to the total observed emission and this is given in the point-like
flux column in Table 3.  It can be seen from the comparison between
the observed flux and the point-like flux that this correction factor
is small, between $<5$ to 10\% on average.

\begin{table*}
  \caption{FWHM, ellipticity  $e$ and position angle (PA) for the 
    standard star observations and galaxies. The measurements marked 
    with $^*$ and $^{**}$ are produced by using centroiding for 
    exposure with an integration time of a few seconds ($^*$) or around 
    one minute ($^{**}$); see Section \ref{reduction} for details.  }
  \label{table_size}
  \begin{tabular}{@{}l|ccccccc|cccccc}
    \hline
                &HD& \multicolumn{6}{c}{PSF star observations} & \multicolumn{6}{c}{Galaxy observations}\\
    Object      & \multicolumn{3}{c}{11.8 $\mu$m} & \multicolumn{3}{c}{18.7 $\mu$m}
                && \multicolumn{3}{c}{11.8 $\mu$m} & \multicolumn{3}{c}{18.7 $\mu$m} \\
                && FWHM & $e$  & PA & FWHM & $e$  & PA  & FWHM & $e$  & PA  & FWHM & $e$  & PA  \\
                && \arcsec &   &\degr&\arcsec&    &\degr&\arcsec&     &\degr&\arcsec&     &\degr \\
    \hline  
    NGC 1097    &~16815&0.35~   & 0.10 & -86 &0.50~~     & 0.15 & -82  &0.37    & 0.13 & 3  &0.54~~     & ...  & -9  \\
    NGC 1386    &~26967&0.33$^*$& 0.21 & -80 &0.53~~     & 0.08 & -70  &0.36$^*$& 0.27 & 1  &0.58~~     & 0.28 & -2  \\
    NGC 1566    &~26967&0.34~   & 0.19 & -79 &0.49~~     & 0.03 & 6    &0.32~   & 0.38 & -75&0.50~~     & 0.56 & 16 \\
    ESO 434-G040&~75691&0.31$^*$& 0.11 & 87  &0.48~~     & 0.10 & -79  &0.31$^*$& 0.10 & -83 &0.50~~     & 0.28 & -63  \\
                &~75691&...     & ...  &...  &0.50~~     & 0.05 & -79  &...     & ...  & ... &0.51~~     & 0.06 & 40 \\
    Mrk 1239    &~83425&0.33~   & 0.15 & -83 &0.49~~     & 0.09 & -69  &0.32~   & 0.09 & -79 &0.49~~     & 0.12 & -83   \\
    NGC 3783    &102964&0.31$^*$& 0.17 & -80 &0.50~~     & 0.08 & 85   &0.32$^*$& 0.02 & 82  &0.50~~     & 0.15 & -13  \\
    M87         &108985&0.31~   & 0.10 & -81 &...        & ...  & ...  &0.35~   & 0.55 & -90 &...         & ...  & ... \\
                &108985&0.32~   & 0.11 & 89  &...        & ...  & ...  &0.28~   & 0.72 & -2  &...         & ...  & ... \\
    NGC 5128    &119193&0.31$^*$& 0.11 & -71 &0.51$^{**}$& 0.45 &  -82 &0.31$^*$& 0.05 & -76  &0.52$^{**}$& 0.11 & 47 \\
    NGC 5506    &124294&0.32$^*$& 0.24 & -78 &0.49$^{**}$& 0.15 &  -83 &0.33$^*$& 0.26 & 86   &0.52$^{**}$& 0.22 & 89  \\
    Circinus    &128068&0.35$^*$& 0.14 & -67 &0.50$^{**}$& 0.08 &  -78 &0.36$^*$& 0.11 & -44  &0.51$^{**}$& 0.17 & -77  \\
    NGC 7469    &~~5112&0.32$^*$& 0.17 & 86  &0.49~~     & 0.14 & 80   &0.33$^*$& 0.51 & -74  &0.51~~     & 0.10 & 33 \\
    NGC 7582    &~~2261&0.33~   & 0.11 & 83  &0.49~~     & 0.12 &  -87 &0.33~   & 0.06 & -45  &0.70~~     & 0.27 & -11  \\
    \hline
  \end{tabular}
\end{table*}

\begin{table*}
  \caption{Photometry: columns \#2 and \#3 are integrated fluxes in a radius $r$, 
in brackets;  errors refer to background noise; \#4 is an upper limit to the 
nuclear size; \#5 and \#6 are nuclear point-like source fluxes, values marked 
with $^f$ are  inferred from the fitting analysis; all others are   from aperture 
photometry (see Sect.\ \ref{sec_nuc}); \#7 \& \#8  are fluxes from the extended 
component, whose size and orientation are  given   in \#9.\label{table_flux}}
  \begin{tabular}{lcccccccl}
    \hline
                 & \multicolumn{2}{c}{Total flux in (radius r)} & Nucleus size &\multicolumn{2}{c}{Unresolved source} & \multicolumn{2}{l}{Circumnuclear} & Extended emission \\
    Object       &  11.8 $\mu$m & 18.7 $\mu$m & 11.8 $\mu$m& 11.8 $\mu$m & 18.7 $\mu$m & 11.8 $\mu$m & 18.7 $\mu$m  &  location and size at radius r \\
                 &  mJy            & mJy              &pc & mJy    & mJy     &mJy & mJy & \\
    \hline
    NGC 1097     & ~25$\pm$2(0.75) & ~41$\pm$4(0.75)  &27 & ~26    & ~49     &        &$>$500  & stellar ring at $ r \goa 8\arcsec$\\
    NGC 1386     & 353$\pm$4(1.13) & 757$\pm$6(1.13)  &32 & 195$^f$& 330$^f$ &170$^f$ &430$^f$ & along ion. cone at $ r \loa 1\arcsec$ \\
    NGC 1566     & ~59$\pm$3(0.75) & 114$\pm$4(0.75)  &31 & ~63    & 128     &        &        & \\
    ESO 434-G040 & 573$\pm$6(1.50) & 1540$\pm$7(1.13) &60 & 590    &1450     &        &        & \\
    Mrk 1239     & 570$\pm$4(1.13) & ~890$\pm$10(1.13)&134& 590    & 930     &        &        &\\
    NGC 3783     & 535$\pm$4(1.13) & 1470$\pm$10(1.13)&69 & 520    &1400     &        &90      &North at r$\loa 2 \arcsec$ (18.7 $\mu$m) \\
    M87          &~16$\pm$0.8(0.75)& $\sim<$18 (0.75)            &35 & ~17    &         &1.6     &        & jet-knot at r$\sim$ 0\farcs87 (HST1)\\
    Sombrero     & $<$7            & ...              &...& ...    & ...     &        &        &\\
    NGC 5128     &1150$\pm$5(1.13) & 2900$\pm$30(1.5) &5  &1150$^f$&2300$^f$ &30$^f$  &770$^f$ & along jet at r $< 1\arcsec $ (18.7 $\mu$m)\\
    NGC 5506     & 958$\pm$5(1.5)  & 1990$\pm$20(1.5) &47 & 900$^f$&1400$^f$ &65$^f$  &590$^f$ &East to West at $r\loa1\farcs5$ \\
    Circinus     &11400(2.63)      &24700(2.63)       &7  &9300$^f$&17600$^f$&2100$^f$&7400$^f$&along ion. cone at  $r\loa2\farcs5$\\
    NGC 7469     & 917$\pm$8(2.25) & 2470$\pm$20(2.25)&99 & 530    &1270     &390     &1200    &stellar-ring at $1\arcsec < r < 2\farcs5$\\
    NGC 7582     & 943$\pm$7(2.63) & 1880$\pm$20(2.63)&29 & 405    & 550     &540     & 1330   &stellar-ring at $ 0\farcs5 <r < 2\arcsec$\\
    \hline
  \end{tabular}
\end{table*}

\section{Description of Galaxies}

\begin{figure*}
  \includegraphics[width=14.4cm]{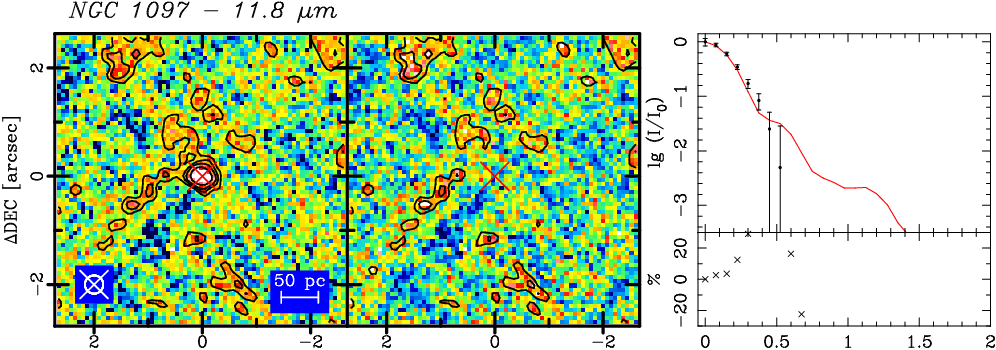}

  \includegraphics[width=14.4cm]{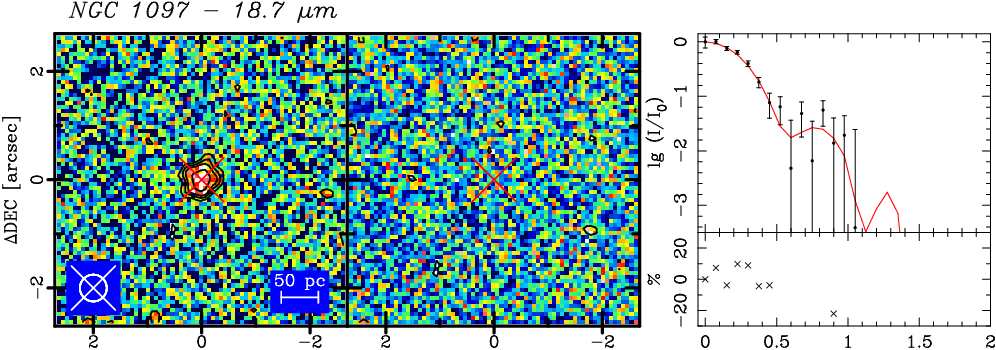}

  \includegraphics[width=14.4cm]{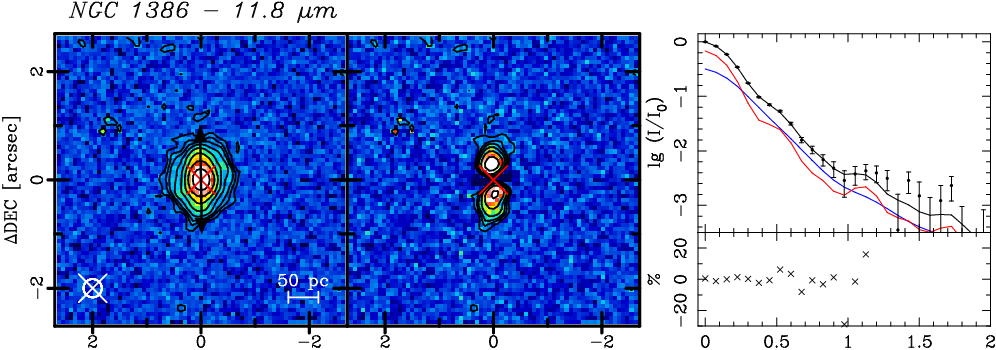}

  \includegraphics[width=14.4cm]{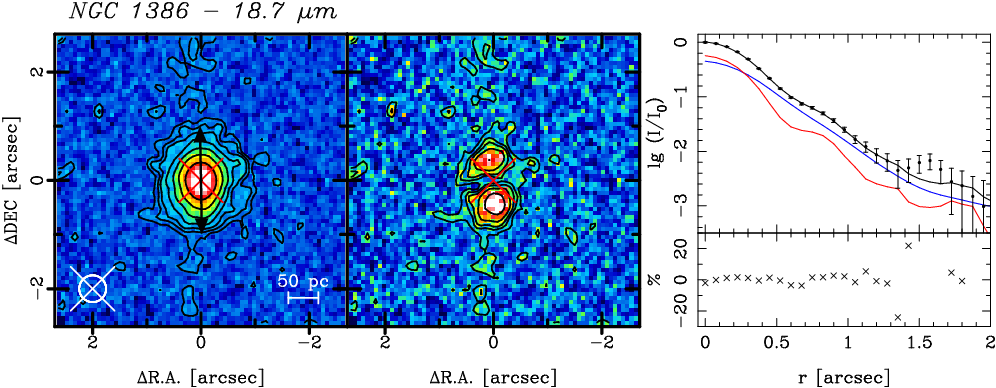}
  \caption
  {The 11.8 and 18.7 $\mu$m images for NGC 1097 and NGC 1386. North is
    up, and east is to the left. \textit{Left}: observed emission, the
    circle on the left corner represents the PSF FWHM, the size of the
    cross at the nucleus indicates the size of the first diffraction
    ring (1.22$\lambda\, D^{-1}$). The location of the ionization cone
    in NGC 1386 is indicated by an arrow.  The 11.8 $\mu$m map of NGC
    1097 suffers from poor sky subtraction.  \textit{Middle}: residual
    after PSF subtraction.  \textit{Upper right}: radial profiles in
    logarithmic flux units normalised at $r=0$ for the galaxy plotted
    with errorbars, the PSF standard (\textit{red} line), fitted disk
    component (\textit{blue} line) and the combined model
    (\textit{black} line).  \textit{Lower right}: the residual
    (observed - model) in percent is shown.  Contours are at
    3$\sigma$, 5, 7, 11, 19 levels; 1 $\sigma$ in NGC 1097 is 29 mJy 
    arcsec$^{-2}$ at 11.8 $\mu$m, 16 mJy arcsec$^{-2}$ at 18.7
    $\mu$m; in NGC 1386 is 19 mJy arcsec$^{-2}$ at 11.8 $\mu$m, 33
    mJy arcsec$^{-2}$ at 18.7 $\mu$m.  \label{fig1097_1386}}
\end{figure*}

Figures \ref{fig1097_1386}, \ref{fig_1566_434}, \ref{fig_1239_3783},
\ref{fig_87_5128}, \ref{fig_5506_circ} and \ref{fig_7469_7582} present
VLT-VISIR  11.8 and 18.7 $\mu$m  images for the sample galaxies. For each 
image, there is  a residual map after a PSF
star subtraction and a  radial profile analysis.  Only the central 
$5\times5$ arcsec$^2$ region is shown,  as no emission is detected further 
out except in the case of NGC 1097 (Fig. 3). In all cases, the lowest
contour level displayed is  3$\sigma$.

A brief description of the results for each galaxy is provided in
following sections. The VISIR nuclear fluxes are compared with the
large-aperture satellite measurements generally used at these mid-IR
wavelengths.  \textit{IRAS} 12 and 25 $\mu$m are taken as a reference.
The \textit{IRAS} 12 $\mu$m bandpass is 8.5--15 $\mu$m and the
\textit{IRAS} 25 $\mu$m bandpass is 19--30 $\mu$m. These bands include
the silicate absorption feature, several PAH features and emission
lines.  On the other hand, the VISIR filters are narrow-band and
mostly line-free. The comparison illustrates the difference between
large-aperture satellite and current inferred AGN fluxes on scales of
FWHM $<\sim0.5$ arcsec resolution.

\begin{figure}
  \includegraphics[height=8cm]{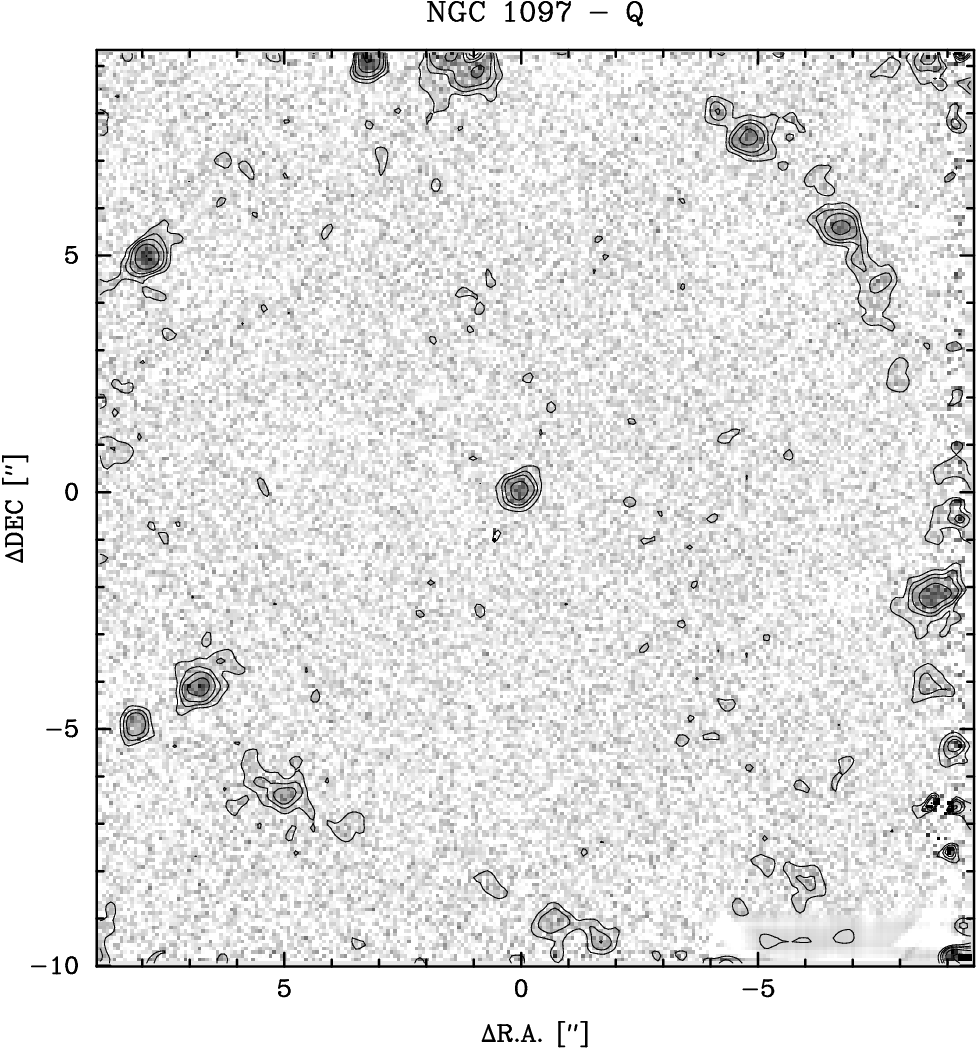}
  \caption{Wide-field VISIR 18.7 $\mu$m image of NGC 1097 showing the
    kpc-radius star-forming ring. Up to more than 12 clumps are
    isolated in this image above the 3 sigma level (first contour in
    the image). 1 $\sigma$ is 16 mJy arcsec$^{-2}$ \label{fig_1097_wide}}
\end{figure}

\subsection{NGC 1097}

NGC 1097 has a LINER/Seyfert 1 nucleus surrounded by a star-forming
ring of radius $r\simeq8$ arcsec or $\sim$600~pc. 
 In the 2--4.5 $\mu$m range, the nuclear region is unresolved 
down to scales of FWHM $<10$ pc from adaptive optics VLT/NaCo images \citep*{Prieto2005}.  

Figure~\ref{fig1097_1386} shows the central 5\,$\times$\,5 arcsec$^2$
region in the 11.8 and 18.7 $\mu$m VISIR images. The most prominent
feature is the nucleus, unresolved at the achieved spatial resolution:
FWHM $<$ 30 pc. The residual images after the subtraction of the
associated PSF star do not reveal any extended nuclear emission and
the radial profiles show that both the nucleus and the PSF star
profiles agree to within the errors.  The complete VISIR field of view
at 18.7 $\mu$m (Fig. 3) is one of the deepest images of the
starforming ring and nucleus of this galaxy at this wavelength,
presenting more than 12 starforming knots in the ring. The VISIR 11.8
$\mu$m image is worse in quality, due to a high background residual,
and only three regions of the ring are detected.  More star forming
regions at this wavelength are detected by \citet{Mason2007} using
Gemini/T-ReCS.

The nuclear fluxes reported by \citet{Mason2007} in a 1.5 arcsec
aperture radius are larger by a factor 2 at 11 $\mu$m and a factor 1.5
at 18 $\mu$m than those found by us, either after integrating in an
aperture containing all the observed emission or from the radial
profile analysis (Table 3). The reason for this difference is not
clear. Mason's et al filters are much broader than the ones used by us
(their filters are 18.3 $\mu$m filter with half bandwidth = 1.51
$\mu$m and 11.7 $\mu$m filter with 1.13 $\mu$m half
bandwidth). Accordingly, their 11.7 $\mu$m may include the
contribution of the usually strong 11.2 $\mu$m PAH feature.  However,
the results from IR nuclear spectroscopy carried by the same authors
indicate the absence of other PAH features in the nucleus of NGC 1097.
Thus, the plausible explanation for the difference is genuine nuclear
variability, by a factor of 2 or so in a time scale of 1 year, the
spanning time between both observation sets.

Comparing with large aperture \textit{IRAS} (Table 1) or
\textit{Spitzer} (6.6 Jy at 24 $\mu$m; \citealt{Dale2007}) fluxes,
they are two orders of magnitudes larger than the nuclear fluxes
reported here.  In this case, the IR satellite fluxes are dominated by
the prominent star-forming ring of NGC~1097. The observed 18.7 $\mu$m
flux in the ring is $\goa$ 500~mJy (no reliable estimate is possible
at 11.8 $\mu$m). This is a lower limit due to the observational
limitation of chopping; also the full extension of the ring is not
included within the VISIR field of view. Still this lower limit is
already an order of magnitude higher than that of the AGN, indicating
that at mid-IR wavelengths star formation and not the AGN is the
dominant contributor. This is full in line with the result derived
from the analysis of the spectral energy distribution of this nucleus
based on very high resolution data, from UV to optical to IR. This
study shows that the genuine AGN luminosity in NGC 1097 is indeed a
tiny fraction, less than 1\%, of the total IR luminosity integrated
over the galaxy \citep{Prieto2009}.

\subsection{NGC 1386}

NGC 1386 is an inclined, barred lenticular galaxy with a Seyfert 2
nucleus.  The nuclear region is crossed by dust lanes, which are
preferentially distributed along the galaxy major axis at
PA$\sim$20\degr. The {\oiii} 5007 {\AA} emission is highly collimated
and extends along the same direction, north-to-south, up to $\sim$3
arcsec radius from the center (Schmitt et al. 2003; Fig. 2).

\begin{figure}
  \includegraphics[height=8cm]{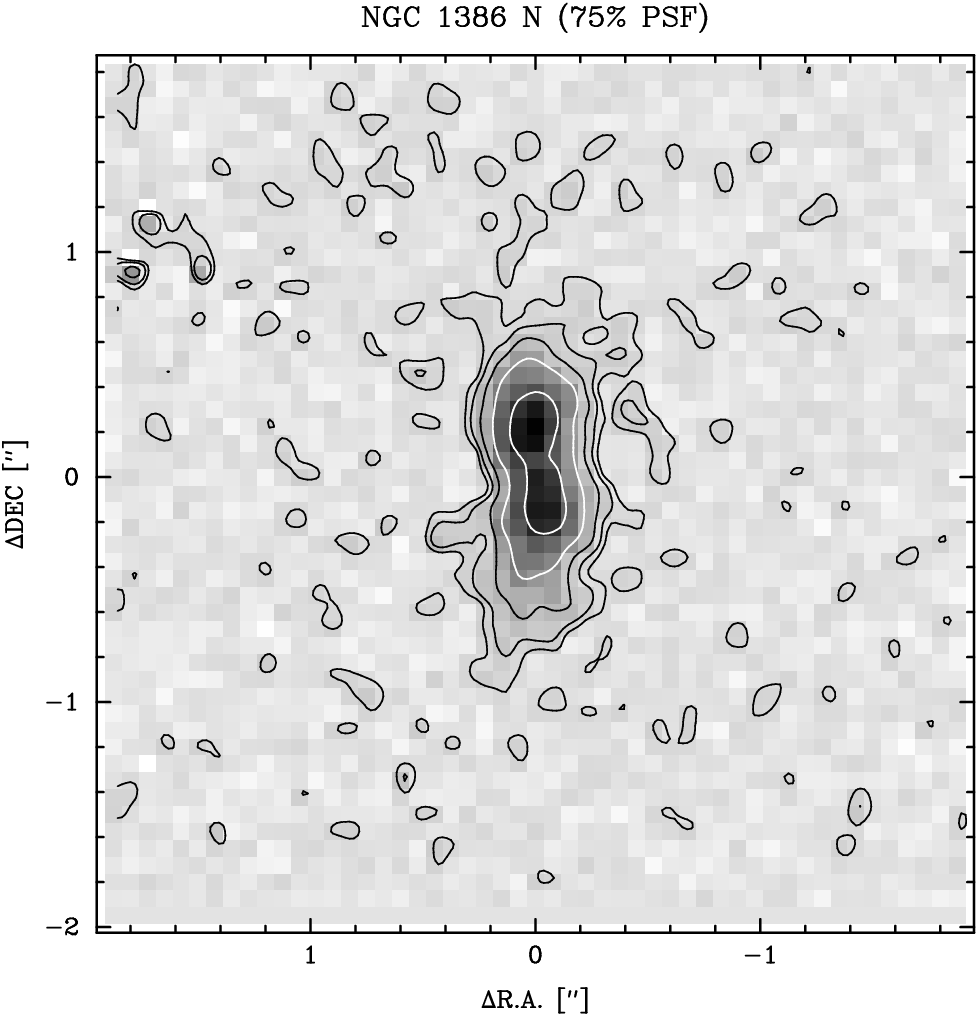}
  \caption{The 11.8 $\mu$m emission of NGC 1386 with the PSF scaled to
    75\% of the peak flux of the nucleus and subtracted. The contours
    are placed at logarithmic intervals at
    1.5, 3, 6, 12 and 24$\sigma$.\label{res_1386_n}}
\end{figure}

Both 11.8 and 18.7 $\mu$m VISIR images also show elongation along the
north--south direction but up to $\sim$1 arcsec from the centre (Fig.\
\ref{fig1097_1386}). Although the elongation is in the direction of
telescope chopping and could be caused by this, we tend to believe
that it is real as it is present at both wavelengths, covering a
similar size. None of the PSF star observations in either band show
similar structure ( see table 2). Also, the symmetric double-lobe
residual seen after subtracting the PSF ( Fig. 2) remains the same
even if the PSF is rotated by 90 degrees.  As north--south is a
preferential emission direction in this galaxy we consider the
extension to be genuine. The radial profiles at both wavelengths
(Fig.\ \ref{fig1097_1386}) are similar, presenting excess emission up
to even larger distances of 1.5--2.0 arcsec.  Figure \ref{res_1386_n}
shows a residual image at 11.8 $\mu$m after scaling the PSF star to
75\% of the galaxy peak. The scale factor is chosen to minimize
negative fluxes seen when scaling to 100 \% of the peak (Fig.\ 2). The
resulting morphology is now smoother and continuous, in line with that
seen in the {\oiii} line emission.

The total integrated flux at 11.8 $\mu$m is 15\% difference from that
reported by \citet{Siebenmorgen2004} using ESO-3.6m / TIMMI2 with a
much broader filter centered at 11.9 $\mu$m (FWHM = 2.26 $\mu$m).
Since this filter includes the strong PAH feature at 11.2 $\mu$m, the
flux difference can partially be attributed to this contribution and
the photometry errors.
 
The results from the profile fitting analysis further indicate that
the contribution of the unresolved source (the AGN) is $\sim$50\% of
the total measured at each of the wavelengths, with the other half
being associated with the north-south extended emission.  Comparing
with the large aperture \textit{IRAS} data, the inferred AGN
contribution is again much less, $\sim$40\% of the \textit{IRAS} 12
$\mu$m and $\sim 20\% $ of the \textit{IRAS} 25 $\mu$m flux levels. It
can be readily seen that the difference with the IRAS 12 $\mu$m flux
is the contribution of the extended nuclear emission seen by
VISIR. This is not the case at 25 $\mu$m, and thus most of the
emission at this wavelength must come, plausibly, from a more diffuse
extended emission over the galaxy.

\begin{figure*}
%5.7
  \includegraphics[width=15.4cm]{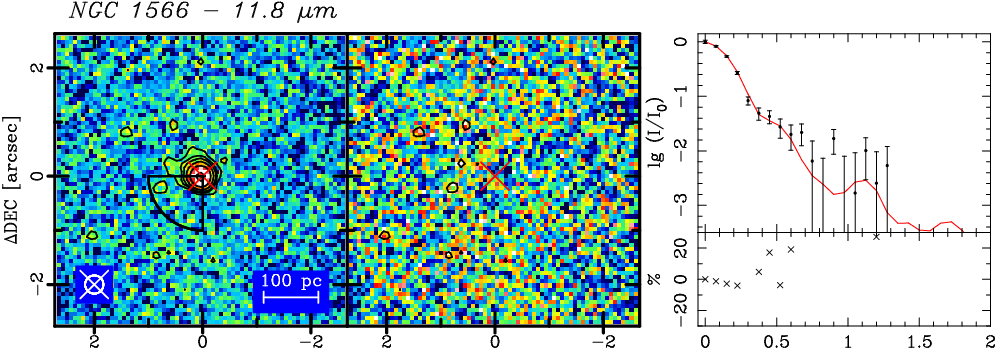}

  \includegraphics[width=15.4cm]{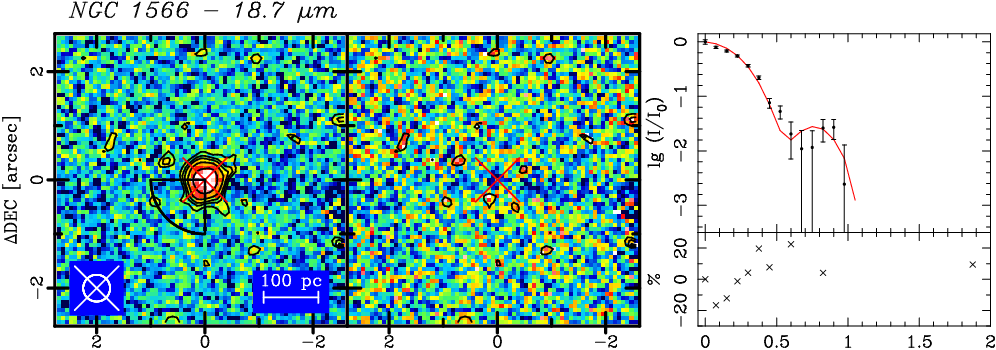}

  \includegraphics[width=15.4cm]{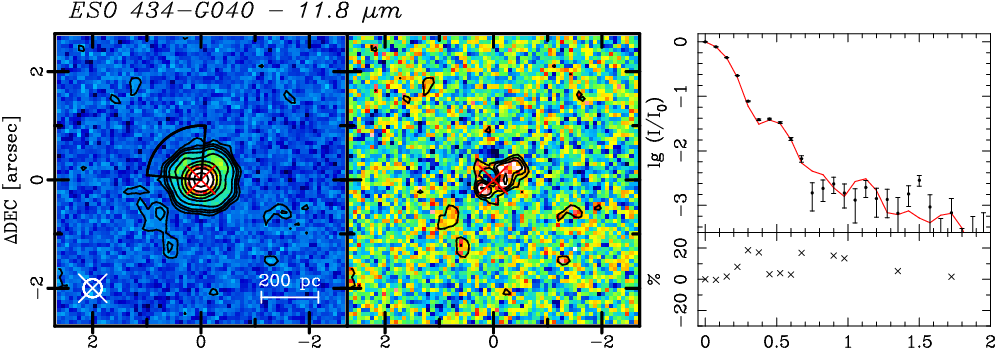}

  \includegraphics[width=15.4cm]{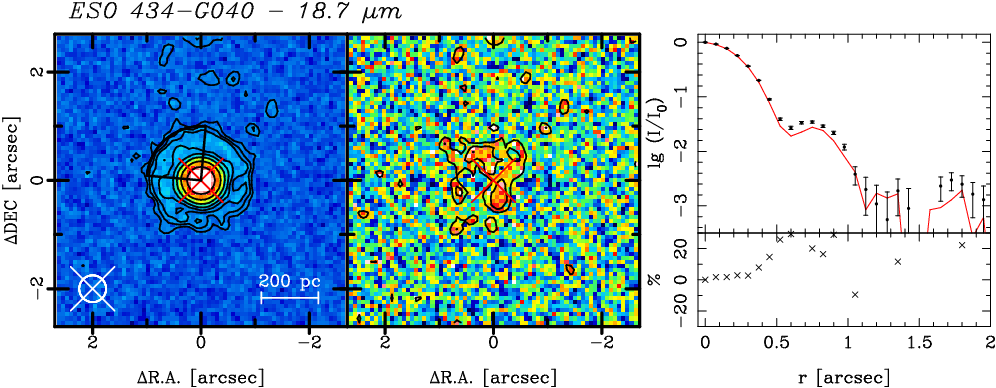}

  \caption{As in Fig.~\ref{fig1097_1386} but for NGC 1566 and
    ESO434-G40. The location of the respective ionised-gas cones is
    outlined in the figures. 1 $\sigma$ level is 22 mJy arcsec$^{-2}$ at 11.8, 42 mJy arcsec$^{-2}$ at 18.7 $\mu$m, for
    NGC 1566; 32 mJy arcsec$^{-2}$ at 11.8, 55 mJy arcsec$^{-2}$
    at 18.7 $\mu$m, for ESO 434-G040.  \label{fig_1566_434}}
\end{figure*}

\subsection{NGC 1566}
NGC~1566 is a face-on spiral galaxy with a Seyfert 1 nucleus.  The
nucleus is known to be variable from X-rays to IR \citep{Baribaud1992,
  Glass2004}. At near-IR wavelengths, 1--4 $\mu$m adaptive optic
images, the nucleus is unresolved down to scales of FWHM $<11$ pc at 2
$\mu$m (Prieto et al. 2007, 2009).  The \textit{HST} {\oiii} images
show a one-sided ionization cone towards the south-west direction
\citep{Schmitt1996}.

The VISIR images (Fig.\ \ref{fig_1566_434}) at both wavelengths just
reveal a dominant central unresolved source with size FWHM $<$ 31 pc
at 11.8 $\mu$m. The \textit{[IRAS} 12 $\mu$m and ISOCAM 8.5--10.7
$\mu$m fluxes \citep{Ramos2007} are larger, by factors of 13 and 4
respectively, than the 11.8 $\mu$m nuclear flux reported here.  the
\textit{IRAS} 25 $\mu$m flux is a factor 10 higher than the 18.7
$\mu$m flux. The smaller difference with Ramos et al. is because of
the deconvolution procedure applied to the ISOCAM radial light
profile, which helps reducing the galaxy contribution, but obviously
is insufficient to overcome the poor ISO spatial resolution.  Thus,
most of the emission measured in this object in IR large aperture data
is dominated form other sources in the galaxy, and the AGN
contribution to the total is minor. The analysis of the SED of this
nucleus based on very high resolution data from UV to optical to IR
shows that indeed that is the case: the intrinsic AGN luminosity is
two orders of magnitude lower than that of the galaxy \citep{Prieto2009}.

\subsection{ESO 434-G040}

ESO~434-G040 (MCG-05-23-016) is an SA galaxy with a Seyfert 2 nucleus
but with  broad Paschen and Brackett series {\hii} lines
\citep{Veilleux1997}. A dust lane crosses the nuclear region on its south-eastern
side, at a distance of less than 1 arcsec from the centre. The 
{\oiii} gas emission extends in the  north-east--south-west direction by 1 arcsec
\citep{Ferruit2000},  as shown in Fig. 5.

The VISIR images are dominated by a bright point-like source
(Fig.\ \ref{fig_1566_434}). The structure seen in the residual image  
at 11.8 $\mu$m is an artefact (sect 3.1); the nature  of the residuals   
at   the 18.7 $\mu$m image is  uncertain, however, there is   marginal excess 
emission in the  18.7 $\mu$m radial profile ($\sim$190~mJy, see Table 3).  
No reported \textit{IRAS} measurements were found.

\begin{figure*}
  % 5.7
  \includegraphics[width=15.6cm]{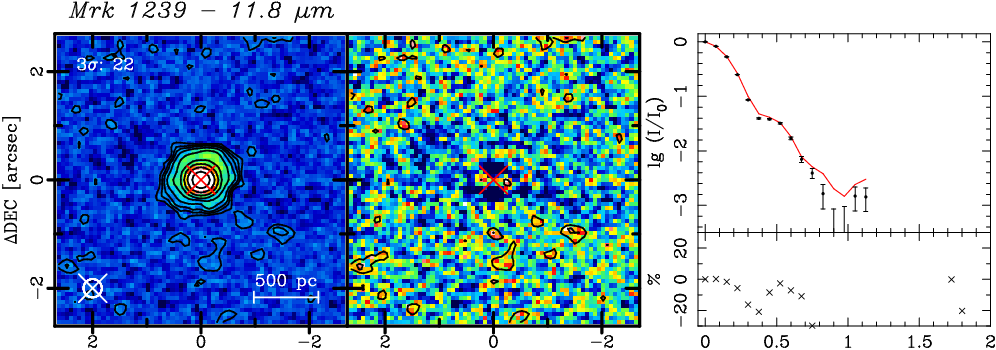}

  \includegraphics[width=15.6cm]{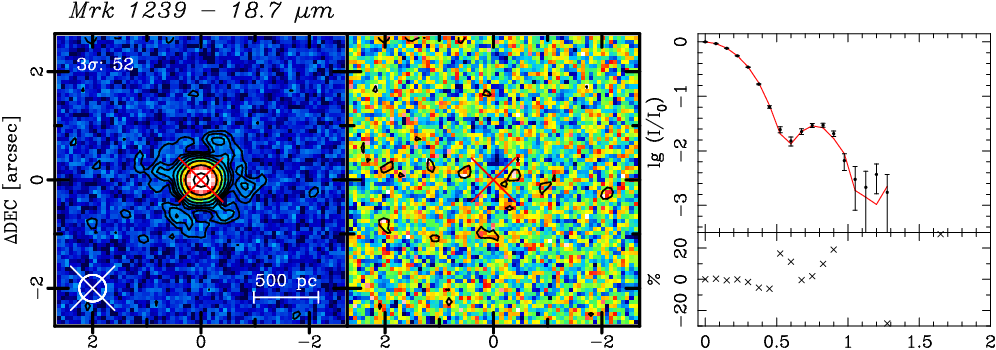}

  \includegraphics[width=15.6cm]{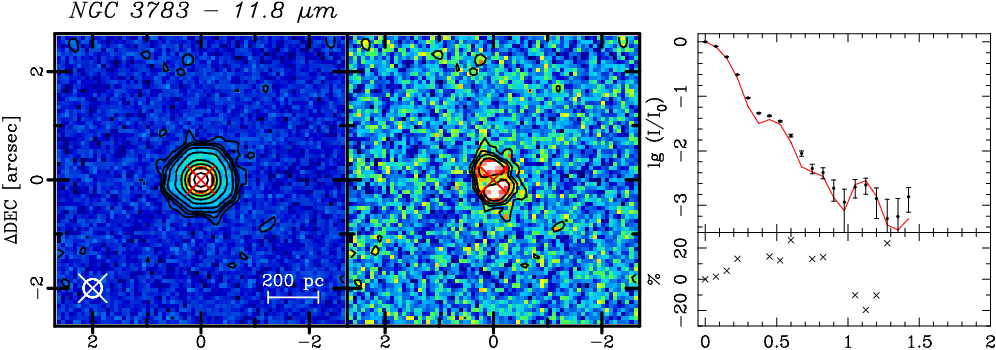}

  \includegraphics[width=15.6cm]{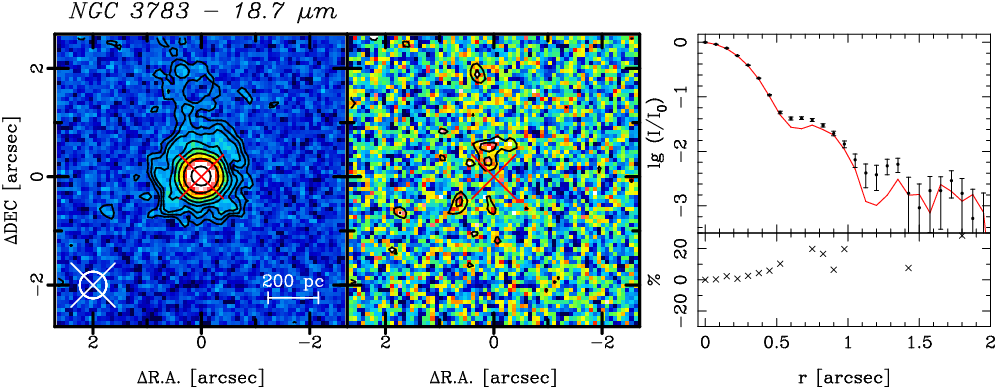}
  \caption{As in Fig.~\ref{fig1097_1386} but for Mrk 1239 and NGC
    3783. 1 $\sigma$ level is 23 mJy arcsec$^{-2}$ at 11.8,
    55 mJy arcsec$^{-2}$ at 18.7 $\mu$m, for Mrk 1239; 23 mJy
    arcsec$^{-2}$ at 11.8, 70 mJy arcsec$^{-2}$ at 18.7 $\mu$m, for
    NGC 3783. \label{fig_1239_3783}}
\end{figure*}

\subsection{Mrk 1239}

Mrk 1239 is an elliptical / S0-class galaxy with a Seyfert 1.5 nucleus.
It is  the most distant object in the sample, 20 times the distance to Circinus (Table 1).

VLT/MIDI interferometric observations in the 8--12 $\mu$m range are
consistent with an unresolved nucleus \citep{Tristram2009}.  VLT/NACO
adaptive optic images in the 1--4 $\mu$m range set an upper limit for
the unresolved nucleus down to scales of FWHM $< 38 pc$ at 2 $\mu$m
\citep{Prieto2009}.  VISIR 11.8 and 18.7 $\mu$m images are also
dominated by an unresolved central source with no extended emission
detected in either the residuals or the radial profile analysis (Fig.\
\ref{fig_1239_3783}).  The extended optical ionised gas in this galaxy
shows neither a particular morphology, it extends around the nucleus
up to about 8 arcsec radius (Mulchaey, Wilson \& Tsvetanov, 1996).

Our photometry at 11.8 $\mu$m is in excellent agreement, 5\%
difference, with that reported by Maiolino et al. (1995), who observed
with a 5.3 arcsec-aperture bolometer at the TMT using a N-broad-band
filter centered at 10.6 $\mu$m.  The agreement indicates that this
nucleus is not variable on scales of at least 11 years, and that not
active star formation is occurring nucleus, otherwise emission from
PAH features should have made a difference at the N-band filter
observations.

Mrk 1239 is one of the three galaxies in the sample -- the other cases
are NGC 5506 and NGC 3783 -- where the AGN accounts for almost all of
the IR emission measured in the entire galaxy: the VISIR fluxes are 80
to 90 \% of the \textit{IRAS} fuxes at 12 and 25 $\mu$m respectively
(Tables 1 and 3).

\subsection{NGC 3783}
NGC 3783 is a nearly face-on spiral galaxy with a Seyfert 1 nucleus.
The {\oiii} emission extends over a surrounding halo up to 200 pc from
the centre \citep{Schmitt2003}; higher-ionization coronal gas
({\fevii}, {\fex}) extends further up to 400 pc from the nucleus in
the north--south direction \citep{Rodriguez-Ardila2006Outflow}.

From VLT/NACO adaptive optics images in the 1--4 $\mu$m range, the
nucleus is unresolved down to scales of FWHM $<$ 22 pc at 2 $\mu$m
\citep{PrietoXian, Prieto2009}.  But first VLTI/MIDI interferometry
results in the 8--12 $\mu$m range point to a resolved nuclear
structure with size of $\sim$60--70 mas, $\sim$15 pc size
\citep{Beckert2008}. However this result requires further confirmation
with other base-line observations.
 
VISIR 11.8 and 18.7 $\mu$m images are dominated by a central bright
source (Fig.\ \ref{fig_1239_3783}). The central two-blob structure in
the 11.8 $\mu$m residual image is due to PSF mismatch (see Sect. 3.1
). The radial profile at both wavelengths shows some marginal excess
emission but that may still be introduced by slight variations in the
PSF star profile.  We consider the VISIR emission unresolved at the
S/N limit of these observations, setting an upper limit for the
nucleus size of FWHM $< 68 pc$ at 11 $\mu$m ( Table 2).
 
Our photometry at 11.8 $\mu$m differs by 20 \% with respect to the
value reported by Haas et al. (2007), using also VISIR but a
narrow-band filter just centered on the PAH feature at 11.2 $\mu$m
(FWHM = 0.6$\mu$m). Thus, the higher flux measured by Haas et al. may
be due to PAH contribution - the photometric errors in both
observations are about 10 \%. If that is the case, some level of
starformation may be present at the nuclear region at scales of less
than 70~pc - the resolution of these observations.  Alternatively, the
difference may be due to variability at very low level: these
observations and that of Haas et al are separated by one year.

Comparing with the \textit{IRAS} flux levels at 12 and 25 $\mu$m, the
VISIR nuclear fluxes are $\sim$60\% lower in either band. This is a
small difference as compared with most of the objects in this sample,
emphasising the dominance of this AGN over the host galaxy light.  The
analysis of the spectral energy distribution of this nucleus on the
basis of high spatial resolution data shows indeed that this AGN
behaves like a quasar dominating the integrated light at any aperture
size in the 1 to 100 $\mu$m range (Prieto et al. 2009).

\begin{figure*}
  % 5.7
  \includegraphics[width=15.9cm]{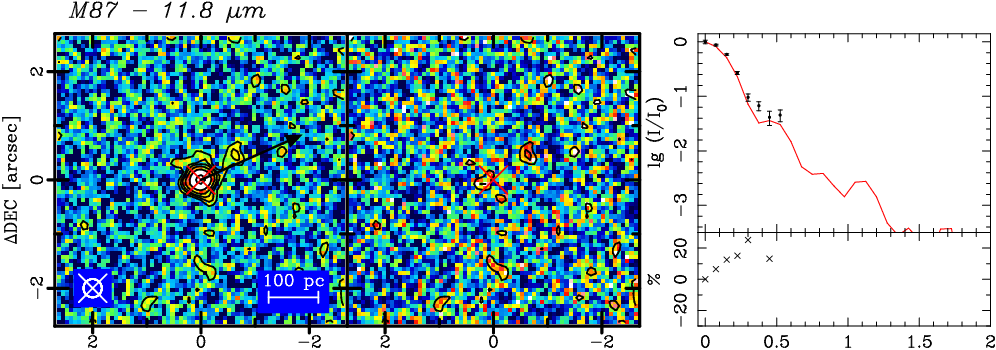}

  \includegraphics[width=15.9cm]{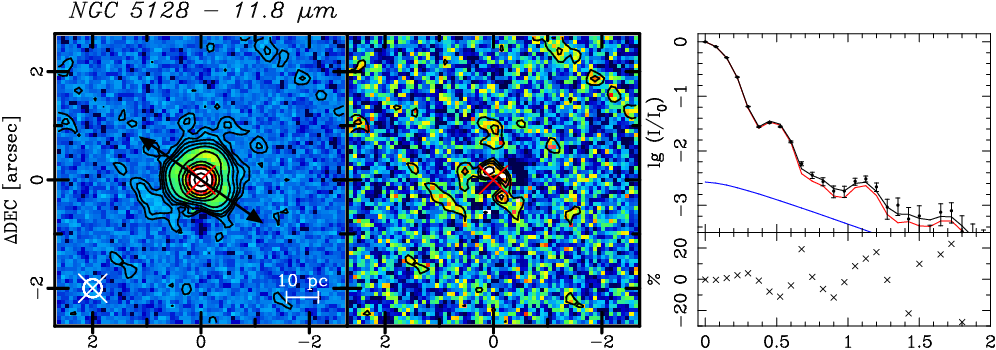}

  \includegraphics[width=15.9cm]{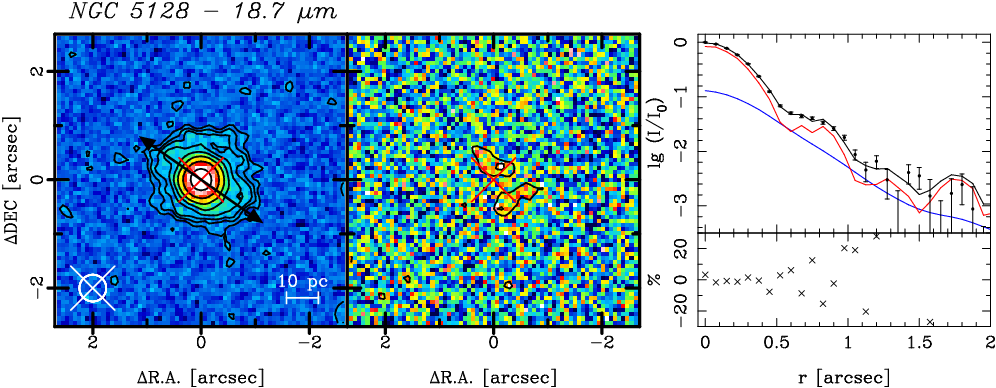}
  \caption{As in Fig.~\ref{fig1097_1386} but for M87 and NGC 5128 (Cen
    A). M87 is barely detected at 18.7 $\mu$m, and it is not
    shown. The diagonal stripes in the 11.8 $\mu$m image of NGC 5128
    are due to sky residuals. The jet directions in M87 and NGC 5128
    are indicated by an arrow in each figure. 1 $\sigma$ level is 8
    mJy arcsec$^{-2}$ at 11.8, 57 mJy arcsec$^{-2}$ at 18.7
    $\mu$m (not shown), for M 87; 32 mJy arcsec$^{-2}$ at 11.8, 123
    mJy arcsec$^{-2}$ at 18.7 $\mu$m, for NGC 5128.
    \label{fig_87_5128}}
\end{figure*}

\subsection{M87}

M87 is a giant elliptical galaxy with a LINER-type nucleus.  It has a
one-sided jet at PA $=-66\degr$ extending up to 2 kpc from the
nucleus. The radio, optical, and X-ray jet has several knots of which
the innermost one -- HST-1 at 0.85 arcsec from the nucleus -- has
brightened fifty-fold in the X-rays in the period 2000--2005
\citep{Harris2006}. A sudden increase in brightness from HST-1 has
also been seen in the IR in our VLT/NACO $K$-band images during the
same period, reaching flux levels comparable to that of the nucleus.

M87 was observed in 2006 with VISIR. At that time, only the nucleus
was bright at 11.8 $\mu$m (Fig.~\ref{fig_87_5128}) but barely detected
at 18.7 $\mu$m.  At 11.8 $\mu$m, a faint source (1.6$\pm$0.3~ mJy;
5$\sigma$) just coinciding with the position of HST-1 is seen at both
the original and the residual images.  A deeper Gemini image by
Perlman et al (2001), using a broad-band filter centered at 10.8
$\mu$m (FWHM= 5.3 $\mu$m) shows additional jet components at larger
distances. They report on marginal nuclear excess emission up to
$r>0.6$ arcsec from the centre. The VISIR image has slightly better
spatial resolution than that by Gemini (0.32 vs. 0.46 arcsec) and so
far our results on the residual image and radial profile analysis are
consistent with the nucleus of M87 being point-like.

The nuclear photometry was done in an aperture radius $r ~< 0.75$
arcsec (Table 3) which avoids the HST-1 region.  The resulting flux at
11.8 $\mu$m is virtually the same as that reported by
\citet{Perlman2001}. Considering the width of the filter used by
Perlman et al. the agreement emphasises the dominance of the continuum
light over any line feature, in particular PAHs, in the spectrum of
M87; also the nucleus was in the same steady state in 2001 (Gemini) as
in 2006 (VISIR), precisely when HST-1 was active.

M87 nucleus is detected at 18.7$\mu$m with VISIR but the signal is
very low, the integrated flux in an aperture including all the
detected emission represents 3 sigma (Table 3).

M87 was not detected in the \textit{IRAS} 25 $\mu$m band. At 11.8 $\mu$m,
the nuclear flux accounts for less than 7\% of the  flux measured in
the \textit{IRAS} 12 $\mu$m band.  Thus, most of the extra emission seen by
\textit{IRAS} has to come from low surface brightness light from the galaxy
itself with a minor contribution from the jet ($\sim$20~mJy is the sum of the jet knots contribution detected in  Perlman et al.).
 
\subsection{The Sombrero Galaxy}

The Sombrero Galaxy (NGC 4594) is a spiral with a Seyfert 1.9/LINER
nucleus. The galaxy is almost edge-on and is crossed mid-plane by a
distinctive dust lane.

It was undetected at 11.8 $\mu$m and thus no observations at 18.7
$\mu$m were attempted.  A formal 6$\sigma$ upper limit for an
unresolved source is 7~mJy.  \citet{Bendo2006} report an 8 $\mu$m flux
of 34.4~mJy based on \textit{Spitzer} imaging; however, such a flux
level would have clearly been detected by VISIR.  Thus, the mid-IR
emission probed by \textit{Spitzer} or \textit{IRAS} has to be largely
diffuse, escaping detection in ground-based observations because of
the chopping technique. The AGN contribution, if any, is tiny at these
wavelengths.

\subsection{NGC 5128}
NGC 5128 is the closest object in this sample. Its jet, at
PA$=55\degr$, is visible from the radio to the X-ray. Adaptive optics
images in the 1--5 $\mu$m range and interferometric spectra in the
8--12 $\mu$m range are both compatible with an unresolved nucleus down
to scales of less than 1 pc \citep{Haering2006, Meisenheimer2007}.

The VISIR images are dominated by a bright central source
(Fig. \ref{fig_87_5128}). At 11.8 $\mu$m the source is rather
symmetric but the radial profile reveals some marginal excess beyond
$\sim$1 arcsec. At 18.7 $\mu$m, the nuclear emission appears elongated
in the direction of the jet. In this case, a fit to the radial
profiles was attempted and the results are shown in the Figure
\ref{fig_87_5128}. At 11.8 $\mu$m, the fit is compatible with a single
point source: the resulting point-like flux and the total integrated
one are the same; at 18.7 $\mu$m, the difference is $\sim$20\%. This
excess represents $\sim600$~mJy and presumably is from the jet.  Our
photometry at 11.8 $\mu$m and 18.7 $\mu$m differ respectively by 50\%
and 20\% with respect to the 11.7 $\mu$m and 17.75 $\mu$m nuclear
fluxes reported by Wysong \& Antonucci (2004). These observations were
done with the Keck telescope in 2002 using filters centered at the
respective wavelengths above indicated, both having a FWHM = 1
$\mu$m. These authors get similar spatial resolutions as the ones
obtained with VISIR, and in both cases the provide flux is integration
over the total detected emission.  The difference in photometry is
thus significant, in particular considering that no PAH features are
seen in the nuclear spectrum (Siebenmorgen et al. 2004).  Cen A's
nucleus is the only source in Prieto's et al study ( 2009) whose high
spatial resolution SED from radio to millimetre to IR can be fit by a
single synchrotron model (Prieto et al. 2007).  Thus, we believe that
the fluxes difference are genuine and linked to the AGN variability.

Comparing with the large aperture \textit{IRAS} 12 $\mu$m and 25
$\mu$m fluxes, the VISIR fluxes are a factor 13 and 8 smaller
respectively.  The analysis of the high spatial resolution SED shows
indeed that the AGN represents a few percent of the total galaxy
emission in the IR.  Thus, most of the \textit{IRAS} emission has to
come from a source other than the AGN, presumably from a low surface
brightness component.

\begin{figure*}
  \includegraphics[width=15.4cm]{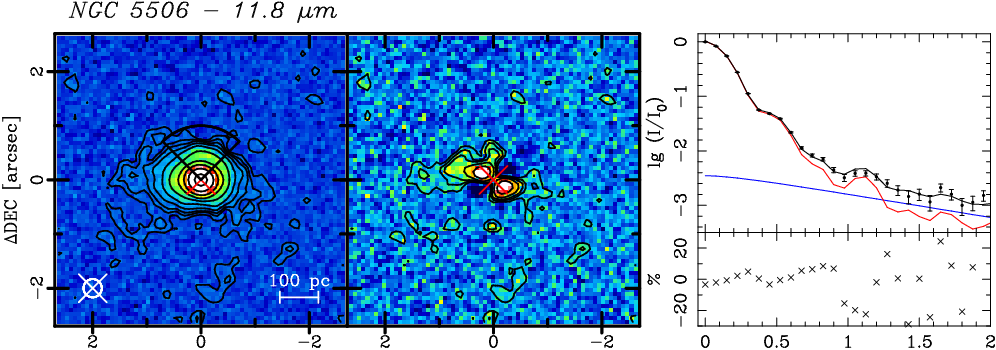}

  \includegraphics[width=15.4cm]{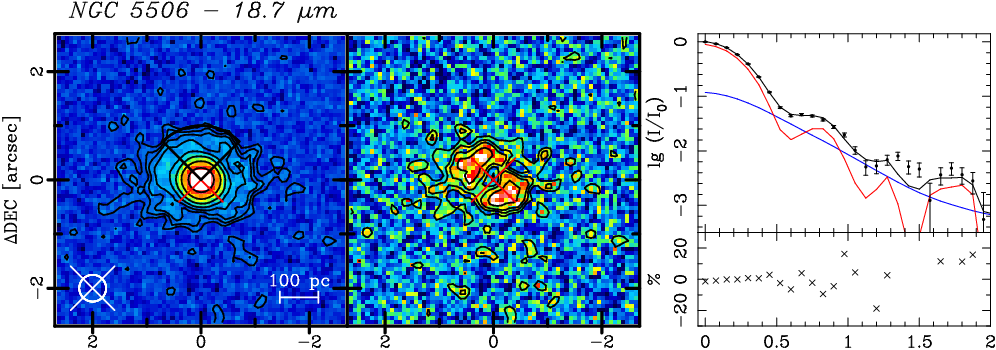}

  \includegraphics[width=15.4cm]{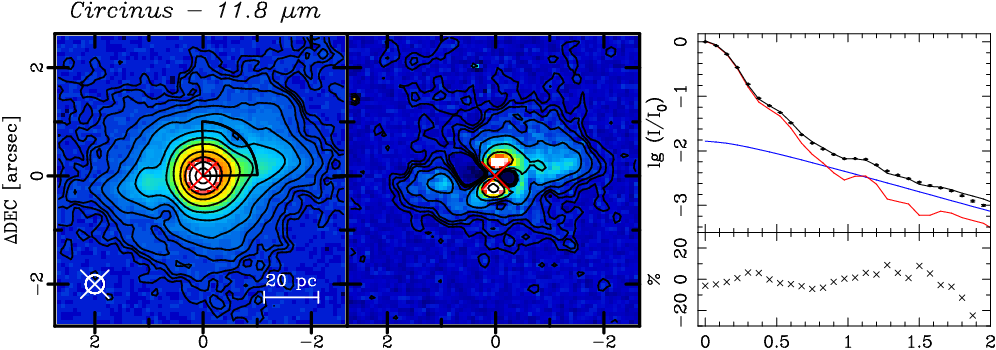}

  \includegraphics[width=15.4cm]{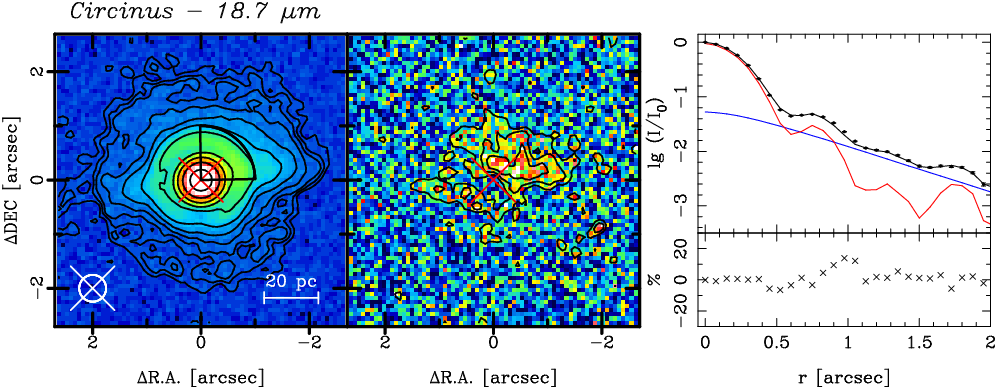}

  \caption{As in Fig.~\ref{fig1097_1386} but for NGC 5506 and
    Circinus.  The direction and morphology of the extended ionized
    gas in Circinus and NGC 5506 are outlined. 1 $\sigma$ level is 23
    mJy arcsec$^{-2}$ at 11.8, 72 mJy arcsec$^{-2}$ at 18.7
    $\mu$m, for NGC 5506; 25 mJy~ arcsec$^{-2}$ at 11.8, 110 mJy
    arcsec$^{-2}$ at 18.7 $\mu$m, for Circinus. \label{fig_5506_circ}
  }
\end{figure*}

\subsection{NGC 5506}

NGC 5506 is an Sa galaxy with a Seyfert type 1.9 nucleus. It is
orientated edge-on and crossed by dust lanes along its mid-plane, in
the east-west direction. Analysis of the HST/ WFPC2-F660W image shows
that the H$\alpha$+[NII] gas extends from the center towards the north
following a cone-like morphology (Fig.~\ref{fig_5506_circ}).

The nucleus, fully obscured at optical wavelengths, reveals in full
realm in VLT-NACO adaptive optics images in the 1--5 $\mu$m
range. These observations set an upper limit to the core size of FWHM
$<10$ pc at 2 $\mu$m \citep{Prieto2009}.

The VISIR 11.8 $\mu$m and 18.7 $\mu$m images are both dominated by the
nuclear source (Fig.~\ref{fig_5506_circ}). Still, there is clear
extended emission, more evident at 11.8 $\mu$m, along the east--west
direction. The residual images are dominated by the symmetric two-blob
structure caused by the PSF mismatch (Sect.\ 3.1), but some weak
emission at radii larger than 1 arcsec is still apparent. The radial
profiles reveal a net excess emission at both wavelengths.

A fit to the radial profiles using a composite model, a PSF star
profile and a disc (Sect.\ 3.2), yields a dominant contribution for
the point-like source at 11.8 $\mu$m, but at 18.7 $\mu$m the
contribution of the extended emission becomes more relevant,
$\sim30\%$ of the total flux (Table 3). Our photometry at 11.8 $\mu$m
agrees very well - $< 5\%$ difference - with that of
\citet{Siebenmorgen2004}, the latter obtained in 2002 with the
ESO-3.6m / TIMMI2 camera, The total integrated fluxes measured in both
cases are compared.

At both 11.8 $\mu$m and 18.7 $\mu$m though, the AGN largely dominates
the IR light of the entire galaxy in a similar situation as it happens
in the previous described AGNs Mrk 1239 and NGC 3783: the unresolved
nuclear fluxes represent 70\%, at 11.8$\mu$m, to 40\% at 18.7 $\mu$m,
of the total flux measured by \textit{IRAS} at 12 $\mu$m and 25
$\mu$m, respectively.  The analysis of the high-spatial-resolution
spectral-energy-distribution of this nucleus shows indeed that this is
another case of a Seyfert nucleus dominating the host galaxy light in
a similar way as a quasar (Prieto et al. 2009).

\subsection{Circinus Galaxy}

\begin{figure}
  \includegraphics[height=8cm]{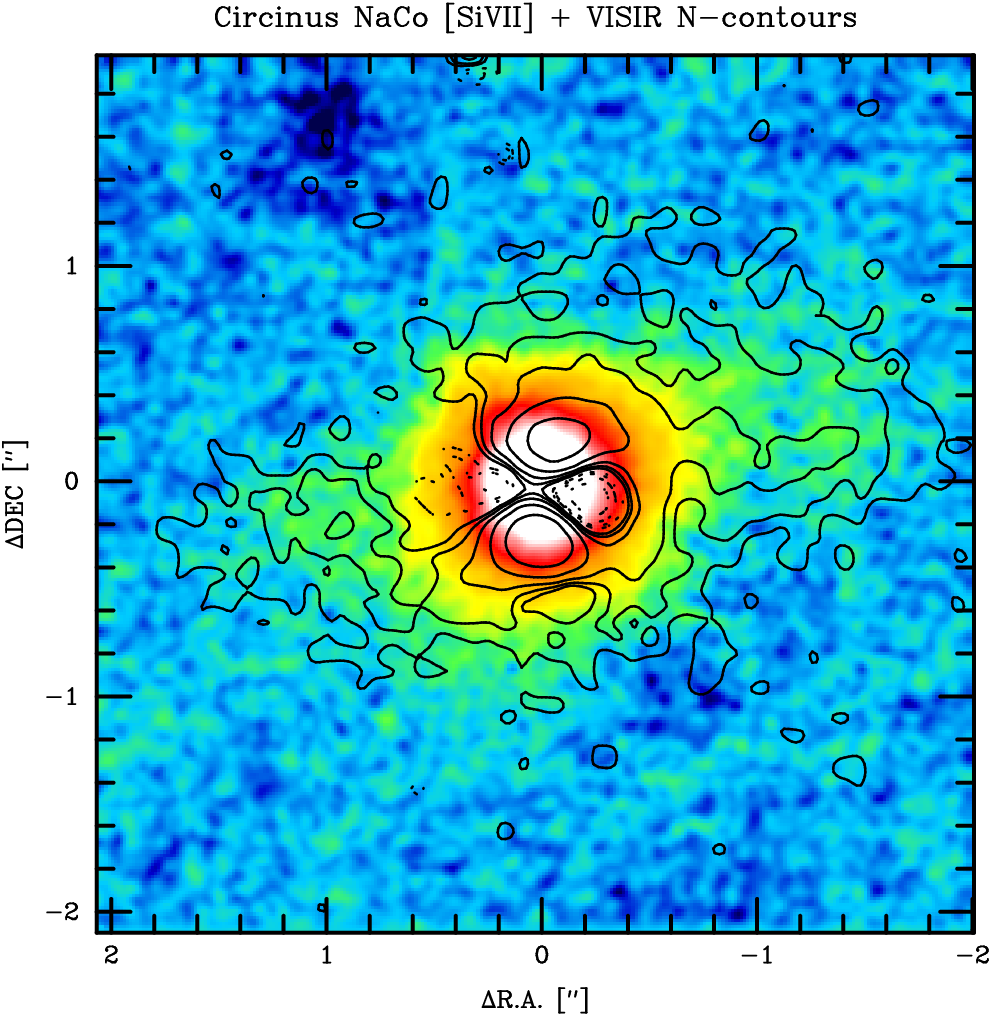}
  \caption{Comparison between the VLT/NaCO [SiVII] 2.48 $\mu$m coronal line emission map (Prieto et al.\ 2005)
    and the VISIR residual image at 11.8 $\mu$m, the latter with 
    contours. The solid contours are at 2.5, 5, 10, 20 and 40$\sigma$
    levels, and dotted contours indicate negative
    residuals. The central two-blob feature in the residual image
is due to PSF mismatch, Sect.\ \ref{PSF}. \label{fig_circ_sivii}}
\end{figure}

The Circinus galaxy and Cen A are the two nearest AGN in this
sample. Circinus is also the brightest in the IR, by an order of
magnitude.  Its optical ionization gas cone \citep{Wilson2000} is
seeing in one direction only, north-west from the nucleus (Fig. 8); in
the IR however, the high ionization coronal gas -- {\sivi} 1.96
$\mu$m, {\sivii} 2.48 $\mu$m -- reveals the counter cone location, at
about the same direction as the optical ionization cone, but more
collimated \citep{Maiolino2000, Prieto2005circ}.

The nucleus of Circinus is resolved at 2 $\mu$m with adaptive optics
observations \citep{Prieto2004, Mueller2006}.  It shows a disc-like
structure, $\sim$2 pc ($\sim0.1$ arcsec) in size, oriented
perpendicular to the ionization cone. The analysis of the
optical-to-IR spectral energy distribution of this structure is
compatible with emission by dust heated at $\sim$300 K by the AGN
(Prieto et al. 2004). Further VLTI / MIDI interferometry in the 8--12
$\mu$m range confirms this structure \citep{Tristrametal2007}.  All
the results together strongly suggests this structure to be the
nuclear torus.

The VISIR images are dominated by a bright central source surrounded
by a large halo ( Fig.\ 7). A similar morphology is seen in the
Gemini/T-ReCS mid-IR images \citet{Packham2005}. Within the halo, a
fairly collimated light beam extends across the nucleus in the
east--west direction. This is better contrasted in the 11.8 $\mu$m
residual image. The beam light is co-spatial with the also rather
collimated emission defined by the {\sivii} 2.48 $\mu$m gas traced
with adaptive optics observations by Prieto et al. (2005), see Fig.\
8.  As this high ionization gas traces pure AGN photons, the common
spatial location is an indication that the mid-IR emission is caused
by dust directly heated by the AGN.

The Circinus nucleus is unresolved in the VISIR images. The achieved
resolution (Table 2) corresponds to a physical scale of FWHM $<7$ pc
at 11.8 $\mu$m. Thus, any further extended emission from the $\sim$ 2
pc scale torus seen at 2 $\mu$m has to extend less than $\sim$ 3 pc
radius from the centre.

A composite point-like source plus disc model (Sect.\ 3.3) was fitted
to the 11.8 and 18.7 $\mu$m radial profiles (Fig.\ 7).  The unresolved
component is the dominant contribution, accounting for $\sim$70--80\%
of the total VISIR flux at those wavelengths.  Deriving the nuclear
fluxes by direct photometry in small apertures lead to similar values.
The VISIR total flux at 18.7 $\mu$m (Table 2) is in very good
agreement with the 18.3 $\mu$m flux reported by Packham et al. (2005),
both measured in a 5 arcsec diameter aperture. The difference is less
than 5\%. That at 11.8 $\mu$m is also in good agreement, 5\%
difference, with the 12 um flux measured in the ESO-3.6m /TIMMI2
spectrum presented in Siebenmorgen et al. (2004). However, Galliano et
al. (2005) reports a total flux with TIMMI2 in the N-broad-band filter
of about 45 \% higher which we do not understand considering these
observations are contemporaneous with those of Siebenmorgen et al.

The contribution of the extended emission component in Circinus is not
negligible in absolute terms, the corresponding fluxes are at the
level of a few janskys at the observed VISIR wavelengths (Table 3).
Comparing with \textit{IRAS} fluxes, the unresolved source at both
VISIR wavelengths accounts for 40 to 20\% of the 12 $\mu$m and 25
$\mu$m IRAS fluxes respectively. As in most objects in the sample, the
IR light of the galaxy dominates over that of the AGN, the relevance
of this contribution becoming higher with increasing wavelength as
revealed by the spectral energy distribution of this nucleus (Prieto
et al. 2009).

\begin{figure*}
  \includegraphics[width=15.4cm]{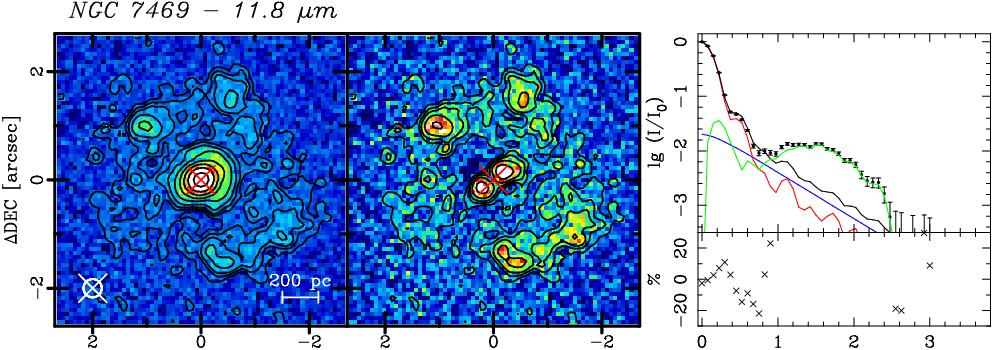}

  \includegraphics[width=15.4cm]{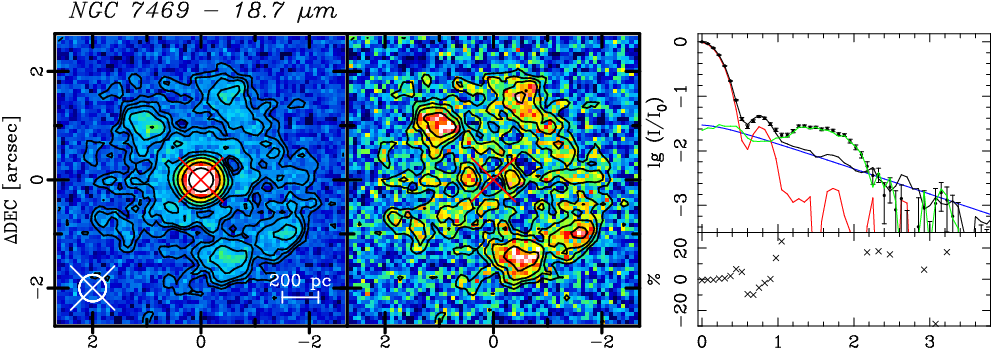}

  \includegraphics[width=15.4cm]{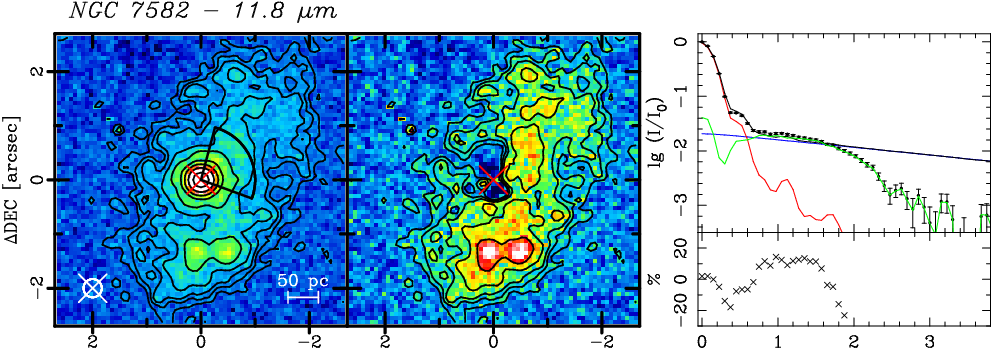}

  \includegraphics[width=15.4cm]{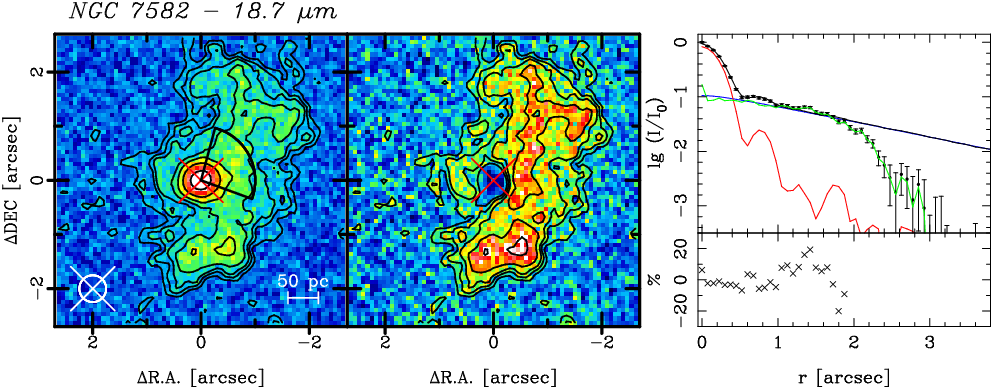}
  \caption{As in Fig.~\ref{fig1097_1386}. The green line is the
    difference between total emission and fitted point-like
    source. The location of the ionization cone in NGC 7582 is
    outlined. 1 $\sigma$ level is 27 mJy arcsec$^{-2}$ at 11.8,
    65$~mJy~ arcsec^{-2}$ at 18.7 $\mu$m, for NGC 7469; 17 mJy~
    arcsec$^{-2}$ at 11.8, 60 mJy~ arcsec$^{-2}$ at 18.7 $\mu$m, for
    NGC 7582. \label{fig_7469_7582}}
\end{figure*}

\subsection{NGC 7469}

NGC 7469 is an SABa galaxy with a Seyfert 1 nucleus surrounded by a
starburst ring at a radius of $\sim$1 arcsec from the center (300 pc).
The current VISIR images at 11.8 and 18.7 are deeper show a very
bright nucleus and several clumps from the star-forming ring (Fig.\
\ref{fig_7469_7582}).  The ring is better contrasted in the residual
images (the central two-blob features in both residual images are
caused by a PSF mismatch, Sect.\ 3.1).

Nuclear fluxes were derived from direct aperture photometry on the
images following the procedure described in Sect.\ \ref{sec_nuc}.  A
composite point-like source plus a disc model (Sect.\ 3.3) fit to the
radial profiles was found not adequate because of the presence of the
star-forming ring.  Nevertheless, despite of this additional
contribution not accounted for in the fit, the nuclear fluxes derived
from the radial profile fit turn to be slightly larger than those from
the aperture photometry.  The total integrated flux - i.e., including
the nucleus and the stellar ring, Table 3 - at 11.88 $\mu$mu differs
by less than 10\% from that by Galliano et al. ( 2005) who uses
ESO-3.6m / TIMMI2 and the broad band filter centred at 11.9 $\mu$m
(FWHM = 2.26 $\mu$m).

The emission from the star-forming ring was estimated as the
difference between the total integrated flux in the image and that of
the nucleus.  This represents about half of the total integrated
emission (Table 3), and some emission from the ring may still be lost
due to the limited nod throw of these observations, 8 arcsec. Thus,
the star-forming ring is at least as powerful as the AGN.  The
corresponding AGN contribution to the IRAS flux is relatively small,
from 40\% at 11.8 $\mu$m to 20\% at 18.7 $\mu$m of the 12 and 25
$\mu$m fluxes respectively. Thus, the remaining IRAS flux at 11.8
$\mu$m most plausibly comes in this galaxy from the star forming ring.

\subsection{NGC 7582}

NGC 7582 is an SB galaxy with a Seyfert type 2 nucleus with broad
recombination lines in the IR \citep{Reunanen2003}.  Several
star-forming regions within a few arcseconds from the nucleus are seen
along the west side of the nucleus \citet{Prieto2002}. These regions
form indeed part of a circumnuclear ring in which more than twenty
independent regions have being isolated in VLT/NACO adaptive optics
images at 2 $\mu$m (Fernandez-Ontiveros et al., in preparation).  The
{\oiii} 5007 {\AA} gas follows a cone-like morphology on the west side
of the nucleus \citep{SB1991}, outlined in Fig.\ \ref{fig_7469_7582}.
The VISIR images show a prominent nucleus, several star-forming
regions from the ring and diffuse emission, the latter being better
contrasted in the residual images. A VISIR image taken with a filter
centered on the {\neii} 12.81 $\mu$m line plus continuum
\citet{Wold2006} reveals a similar morphology.

As in NGC 7469, the nuclear fluxes are also here derived from direct
aperture photometry on the images following the procedure described in
Section \ref{sec_nuc}. The composite PSF plus a disc model yields a
poor fit at radii $r\simeq0.4$ arcsec (Fig.\ \ref{fig_7469_7582}). We
note, however, that the nuclear fluxes derived from a formal fit to
the radial profile differ by $\sim5\%$ at 11.8 $\mu$m and $\sim$9\% at
18.7 $\mu$m, with respect to those derived from the aperture
photometry.  The emission from the star-forming ring was then
estimated as the difference between the total integrated flux in the
image and that of the nucleus (Table 3).

As in NGC~7469, the contribution from the star-formation ring is as
important as that of the AGN. In particular, the emission at 18.7
$\mu$m is almost a factor 3 larger than that of the AGN.  As compared
with \textit{IRAS}, the AGN represents $\sim$30\% of the 12 $\mu$m
\textit{IRAS} flux and $<10\%$ of 25 $\mu$m \textit{IRAS} flux.

%%%%%%%%%%%%%%%%%%%%%%%%%%%%%%%%%%%%%%%%%%%%%%%%%%%%%%%%%%%%%%%%%%%%%%%%%
\section{Discussion and conclusions}

The mid-IR observations presented in this work, reaching spatial
resolutions down to FWHM $\sim$ 0.3 arcsecs, allow us to constrain the
AGN emission at mid-IR wavelengths within regions of 35 pc size in
diameter on average (range form 5 to 130 pc).  Within the central kpc
region, most of the emission is concentrated in the nuclear region,
and in most cases the bulk of it is linked to an unresolved component.
Considering the physical scales sampled in these AGN, the outer radius
of the torus at mid-IR wavelengths should on average be less than 18
pc, less than 4 pc in Circinus, this being the only galaxy in the
sample with a detected parsec scale disc-like structure at its centre
at near- and mid-IR wavelengths. Resolved or extended emission is
detected in most of the objects within a few arcsec of the centre. The
measured contribution of this extended component is a lower limit as
part of the emission, particularly if diffuse, may easily be
subtracted out in gound-based chopped observations.  Further detection
of emission from the galaxy is severely limited by the nod-throw of
these observations, typically in the range of 8--10 arcsec.

On the basis of thirteen  AGN studied, the following results are found:
\begin{enumerate}
\item In three AGN (NGC 7582, NGC 7469 and NGC 1097), strong
  circumnuclear star-forming regions within a few arcsec from the
  centre are detected.  These are located at radii of $\sim$130 pc in
  NGC 7582, 600 pc in NGC 1097 and 900 pc in NGC 7469. In all cases,
  their associated emission is comparable with, or even larger than,
  that of the AGN, particularly at 18.7 $\mu$m.

\item In six further AGN, extended emission on scales from 1 to 3
  arcsec from the centre is detected. This emission is preferentially
  distributed along a particular direction, usually coinciding with
  the ionization cone or the jet direction. In all cases, this
  contribution represents a few per cent of the AGN flux.
 
\item Only in four AGN the emission is concentrated into a central
  unresolved source, and in one galaxy, the Sombrero, no detection is
  reached.

\item In comparing the present photometry with previous works, most
  centered at $\sim$ 11 $\mu$m and using broad filters that include
  PAH features, differences up to 15 to 20\% in some cases which may
  be attributed to the contribution of PAH features are found; in M87
  and NGC 5507 the agreement is within 5\% indicating the absence of
  PHA features in these nuclei and hence of active star formation.
  Only in NGC 1097 and Cen A, whose nuclear spectrum does not present
  PAH features, genuine variability by a factor of 2 at 11.8 $\mu$m in
  a time scale of 1 to 4 years respectively is indicated.
\end{enumerate}

In the six AGN with extended emission about the centre, that appears
tightly collimated.  In Cen~A and M87, it is aligned with the jet (in
M87, it coincides with the brightest jet knot, HST-1). In Circinus,
NGC 1386, and NGC 3783 the extended emission is co-spatial with the
extended ionized gas. Only in NGC 5506, the extended component spreads
along the mid-plane of the galaxy being perpendicular to the optical
ionization cone main axis.  The mid-IR filters used in this work,
narrow and centred on windows where emission lines or PAH features are
unimportant, are thus targeted to measure pure continuum
emission. Thus, the extended emission in M87 and Cen A is presumably
of synchrotron origin; in all other cases, its nature is more
ambiguous. Possibilities include free--free emission due to cooling of
the ionized gas in the ionization cone \citep{Contini2004}, or/and
emission from dust heated by the AGN. Considering the intrinsic AGN
luminosities of these objects, inferred from either high spatial
resolution IR spectral energy distribution \citep{PrietoXian,
  Prieto2009}: $\sim 8\,\times\,10^{42}$ erg/sec in Circinus; $\sim
4\,\times\,10^{44}$ erg/sec in NGC 3783, or X-ray data; $\sim
1.3\,\times\,10^{42}$ erg/sec in NGC 1386 \citep{Levenson2006}, and
the distances at which mid-IR emission is detected -- in the range of
30 pc in Circinus, to 400 pc in NGC 3783 (see Table 3) -- the nature
of this emission could be reconciled with dust directly heated by the
AGN provided its equilibrium temperature is in the 100 K range
(Barvainis's 1987 formalism is assumed).

The contribution of free-free emission in the mid-IR could also be
important if strong shocks exciting the gas are occurring (see, for
example, fig. 2c of Contini \& Viegas-Aldrovandi 1990). Evidence for
high gas velocities in the above objects is inferred from the
kinematic analysis of their high ionization coronal
lines. Specifically, the FWHM of [FeVII] 6087 A is $\sim$400 km
s$^{-1}$ in Circinus and $\sim$1400 km s$^{-1}$ in NGC 1386 and NGC
3783 \citep{Rodriguez-Ardila2006Outflow}. Moreover, the size of the
extended mid-IR emission closely coincides with the observed sizes of
the Fe or Si coronal region (Rodriguez-Ardila et al.) and with their
spatial location, which indicates that a fraction of the mid-IR
emission is due to free--free emission from predominantly shock-heated
coronal gas. Estimating this contribution requires detailed modelling
and is currently being explored.

Mid- and far-IR emission of galaxies is usually derived from
large-aperture data obtained by IR satellites. For galaxies with an
AGN, it is widely assumed that most of this emission comes from the
nucleus.  The flux level of the nuclear point source in the AGN
studied in this work proves the assumption to be inadequate in most
cases. The estimated 11.8 and 18.7 $\mu$m nuclear fluxes are larger,
by factors of $>3$, than the fluxes measured by \textit{IRAS} at 12
and 25 $\mu$m in 70\% of the sample, the largest discrepancy being
more than an order of magnitude in five galaxies: NGC 1097, NGC 1566,
M87, Cen A and NGC 7582 (in the latter case only at 18.7 $\mu$m). In
the remaining 30\% of the sample, the \textit{IRAS} flux levels are
still larger but within a factor of 2.

The ``extra'' IR excess measured by \textit{IRAS} has to come from a
source other than the AGN, either strong IR sources, presumably
located outside the central 20 arcsec $\times$ 20 arcsec region --
common f.o.v.\ in ground-based mid-IR observations -- or from a more
extended low surface brightness component across the galaxy. For
example, in the three AGN with circumnuclear star-forming regions, the
total integrated emission (AGN plus star-forming ring) accounts for
the IRAS flux levels within a factor of 2. This is the case of NGC
7582 and NGC 7469; in NGC 1097, the difference is still a factor 10,
but in this case the VISIR f.o.v.\ does not map the complete extension
of the ring. In all other AGN with detected nuclear extended emission,
this contribution is minor, albeit a lower limit because of the lower
sensitivity imposed by chopped observations. The overall conclusion
based on this small sample is that in AGN with strong circumnuclear
star formation, this component can well account for most of the IRAS
flux. For all other cases, either a low surface brightness component
extending over galactic scales or strong extranuclear IR sources --
e.g. {\hii} regions in spiral arms -- are the major contribution to
the IR light of these galaxies. In either case, the contribution of
these components surpasses by large factors to orders of magnitude
that of the AGN at mid-IR wavelengths.

%%%%%%%%%%%%%%%%%%%%%%%%%%%%%%%%%%%%%%%%%%%%%%%%%%%%%%%%%%%%%%%%%%%%%%%%%
\section{Acknowledgements}
JR acknowledges financial support from the Academy of Finland (projects
8121122, 8127055).

\label{lastpage}

\end{document}